\documentclass[12pt]{article}
\usepackage{graphicx}
\usepackage{amsfonts}
\usepackage{amsmath}
\usepackage{mathrsfs}
\usepackage{setspace}
\def\beq{\begin{equation}}
\def\eeq{\end{equation}}
\def\bea{\begin{eqnarray}}
\def\eea{\end{eqnarray}}
\def\no{\noindent}
\def\nn{\nonumber}
\def\bib{\bibitem}
\def\ba{\begin{array}}
\def\ea{\end{array}}
\def\al{\alpha}
\def\be{\beta}
\def\de{\delta}
\def\tq{\tilde q}

\def\da{\dagger}
\def\l{\langle}
\def\r{\rangle}

\begin{document}

\begin{center}
{\large \bf \sf
Low energy properties of the $SU(m|n)$ supersymmetric \\
Haldane-Shastry spin chain}

\vspace{1cm}
{\sf B. Basu-Mallick$^1$ \footnote{ Corresponding Author: 
e-mail: bireswar.basumallick@saha.ac.in, Phone: +91-33-2337-5346,
FAX: +91-33-2337-4637},
Nilanjan Bondyopadhaya$^1$
\footnote{e-mail: nilanjan.bondyopadhaya@saha.ac.in},
and Diptiman Sen$^2$ \footnote{e-mail: diptiman@cts.iisc.ernet.in}}
\bigskip

{\em $^1$Theory Group, Saha Institute of Nuclear Physics, \\
1/AF Bidhan Nagar, Kolkata 700 064, India}
\bigskip

{\em $^2$Centre for High Energy Physics, Indian Institute of Science, \\
Bangalore 560 012, India}
\end{center}
\bigskip

\vspace {1.4 cm}
\baselineskip=18pt
\no{\bf Abstract }

The ground state and low energy excitations of the $SU(m|n)$ supersymmetric 
Haldane-Shastry spin chain are analyzed. In the thermodynamic limit, it is 
found that the ground state degeneracy is finite only for the $SU(m|0)$ and 
$SU(m|1)$ spin chains, while the dispersion relation for the low energy and 
low momentum excitations is linear for all values of $m$ and $n$. We show 
that the low energy excitations of the $SU(m|1)$ spin chain are described 
by a conformal field theory of $m$ non-interacting Dirac fermions which have 
only positive energies; the central charge of this theory is $m/2$. Finally, 
for $n\ge 1$, the partition functions of the $SU(m|n)$ Haldane-Shastry spin 
chain and the $SU(m|n)$ Polychronakos spin chain are shown to be related in 
a simple way in the thermodynamic limit at low temperatures.

\baselineskip=16pt
\vspace{.8 cm}
\no PACS No.: 02.30.Ik, 75.10.Jm, 05.30.-d, 03.65.Fd 

\vspace {.8 cm}
\no Keywords: Haldane-Shastry spin chain, partition function, low energy 
excitations, conformal field theory

\newpage 

\baselineskip=18pt
\no \section{Introduction }
\renewcommand{\theequation}{1.{\arabic{equation}}}
\setcounter{equation}{0}
\medskip

The Haldane-Shastry (HS) spin-1/2 chain is an integrable model in which 
equally spaced spins on a circle interact with each other through pairwise 
interactions which are inversely proportional to the square of their chord 
distances \cite{hald1,shas}. Interestingly, the HS spin-1/2 chain is easier to
study in some respects than the integrable spin-1/2 chain with nearest-neighbor
interactions; it has a Yangian quantum group symmetry, and the low energy
excitations form an ideal gas with fractional statistics \cite{hald2,ha,poly1}.
The HS spin-1/2 chain has an $SU(2)$ symmetry, with the spin at each site 
forming the fundamental representation of $SU(2)$. This model can be 
generalized to an $SU(m)$ symmetric model whose Hamiltonian is given, with 
$N$ lattice sites, by
\beq H_{HS} ~=~ \frac{1}{2} ~\sum_{1 \le j <k \le N} ~ \frac{1+P_{jk}}{\sin^2
(\xi_j - \xi_k )}, \label{a1} \eeq
where $\xi_j = j \pi /N$, and $P_{jk}$ is the exchange operator which 
interchanges the `spins' (which can take $m$ possible values) on the $j$-th 
and $k$-th lattice sites.

One can find the complete energy spectrum of the HS spin chain in (\ref{a1}),
including the degeneracy of each energy level, through the motif 
representations of its $Y(gl_m)$ Yangian symmetry \cite{hald3,bern,hika1}. By 
using this information about the spectrum, it is possible in principle to 
construct the partition function of this spin chain. However, there is a 
simpler method for calculating the partition function of the $SU(m)$ HS spin 
chain \cite{fink} which uses the so-called freezing technique 
\cite{poly2,suth1,poly3}. The freezing technique consists of taking the strong 
coupling limit of the spin Calogero-Sutherland (CS) Hamiltonian; then the
coordinates of the 
particles freeze at the classical equilibrium positions of the scalar part of 
the potential, and the spins get decoupled from the coordinate degrees of 
freedom. As a result, one can derive the partition function of the HS spin 
chain by `modding out' the partition function of the spinless CS model from 
that of the spin CS model.

There exists an $SU(m|n)$ supersymmetric extension of the HS spin chain 
\cite{hald2}, where each site is occupied by either one of $m$ type of bosonic
states or one of $n$ type of fermionic states. Such supersymmetric spin chains
play a role in describing some correlated systems in condensed matter physics, 
where holes moving in the dynamical background of spins behave as bosons, and 
spin-1/2 electrons behave as fermions \cite{schl,arik,thom}. The $SU(m|n)$ 
supersymmetric HS spin chain exhibits the $Y(gl_{(m|n)})$ super-Yangian 
symmetry \cite{hald2}; this is also the quantum group symmetry of the 
$SU(m|n)$ supersymmetric Polychronakos spin chain. So it is expected that the
spectra and partition functions of these two spin chains would share some 
common features. The freezing technique has been used to compute the partition
function of the $SU(m|n)$ Polychronakos spin chain \cite{basu1,hika2}, after
mapping the corresponding supersymmetric exchange operators to a representation
of the permutation algebra containing `anyon like' spin dependent interactions
\cite{basu2,basu3}. In Ref. \cite{basu4}, this technique has been used to 
compute the exact partition function of the $SU(m|n)$ HS spin chain. 
Subsequently, it was shown that the partition function of the $SU(m|n)$ HS spin
chain can be expressed through the Schur polynomials associated with the motif
representations, and an exact duality relation has been established between the
partition functions of the $SU(m|n)$ and $SU(n|m)$ HS spin chains \cite{basu5}.

In this paper, our main aim is to study the low energy spectrum of the 
$SU(m|n)$ HS spin chain in the thermodynamic limit $N \to \infty$. To this 
end, in Sec. 2 we review some of the results known for this model. We 
subsequently use its exact partition function to compute the complete 
spectrum for finite values of $N$. In particular, we give explicit expressions
for the degeneracies of all the energy levels by taking a limit of the Schur 
polynomials corresponding to the motif representations. In Sec. 3, we 
discuss the momentum eigenvalues associated with the motifs of the $SU(m|n)$ 
HS spin chain. In Sec. 4, we focus on the ground state and low energy 
excitations of the $SU(m|n)$ HS spin chain for all possible values of $m$ 
and $n$. In particular, we study the degeneracy of the ground state and the 
relation between the energy and momentum of the low energy excitations in the
thermodynamic limit. In Sec. 5, we discuss whether the low energy excitations 
can be described by conformal field theories \cite{itzy} for certain values of
$m$ and $n$. In Sec. 6, we explicitly prove the equivalence at low temperatures
of the partition functions of the $SU(m|1)$ HS spin chain and a model of $m$ 
non-interacting fermions with a particular kind of energy dispersion. We also 
derive an interesting relation between the partition functions of the 
$SU(m|n)$ HS spin chain and the $SU(m|n)$ Polychronakos spin chain at low 
temperatures, for any value of $n\ge 1$. We summarize our results in Sec. 7.

\no \section{Energy spectrum of the $SU(m|n)$ HS spin chain}
\renewcommand{\theequation}{2.{\arabic{equation}}}
\setcounter{equation}{0}

For the purpose of defining the Hamiltonian of the $SU(m|n)$ supersymmetric HS
spin chain, let us consider operators like $C_{j \al}^\da$ ($C_{j \al}$) which
create (annihilate) a particle of species $\al$ on the $j$-th lattice site.
These creation (annihilation) operators are assumed to be bosonic when $\al \in 
\{ 1,2,\dots ,m \}$, and fermionic when $\al \in \{ m+1,m+2,\dots ,m+n \}$.
Thus, the parity of $C_{j \al}^\da$ ($C_{j \al}$) is defined as 
\bea && p(C_{j \al}) ~=~ p(C_{j \al}^\da) ~=~ 0 ~~\mathnormal{for}~~ \al \in 
\{ 1,2,\dots, m \}, \nn \\ 
{\rm and} && p(C_{j \al}) ~=~ p(C_{j \al}^\da) ~=~ 1 ~~\mathnormal{for}~~ \al 
\in \{ m+1,m+2,\dots, m+n \}. \eea
These operators satisfy the commutation (anticommutation) relations
\beq [C_{j \al},C_{k \be}]_{\pm} ~=~ 0, ~~ [C_{j \al}^\da, C_{k \be}^\da
]_{\pm} ~=~ 0, ~~ [C_{j \al}, C_{k \be}^\da]_{\pm} ~=~ \de_{jk} \de_{\al \be},
\label{b2} \eeq
where $[A,B]_{\pm} \equiv AB- (-1)^{p(A)p(B)}BA$.
We now consider a subspace of the related Fock space
in which the number of particles on each site is exactly 1, namely,
\beq \sum_{\al=1}^{m+n} ~C_{j\al}^{\da} C_{j\al} ~=~ 1 \label{b3} \eeq
for all $j$. On this subspace, we define the supersymmetric exchange operators
\beq \hat{P}_{jk}^{(m|n)} ~\equiv ~\sum_{\al,\be=1}^{m+n} ~C_{j \al}^\da
C_{k \be}^\da C_{j \be} C_{k \al}, \eeq
where $1 \le j <k \le N$. These $\hat{P}_{jk}^{(m|n)}$'s yield a realization 
of the permutation algebra given by
\beq \mathcal{P}_{jk}^2 =1, ~~\mathcal{P}_{jk}\mathcal{P}_{kl}= 
\mathcal{P}_{jl} \mathcal{P}_{jk} = \mathcal{P}_{kl} \mathcal{P}_{jl}, ~~
[\mathcal{P}_{jk}, \mathcal{P}_{lm}] =0, \label{b5} \eeq
where $j,~k,~l,~m$ are all distinct indices.
Replacing $P_{jk}$ by $\hat{P}_{jk}^{(m|n)}$ in Eq. (\ref{a1}), we obtain 
the Hamiltonian of the $SU(m|n)$ supersymmetric HS model as \cite{hald2}
\beq \mathcal{H}^{(m|n)}_{HS} ~=~ \frac{1}{2} ~\sum_{1 \le j<k \le N} ~
\frac{1+\hat{P}_{jk}^{(m|n)}}{\sin^2(\xi_j-\xi_k)}. \label{b6} \eeq

As shown in Ref. \cite{basu4}, the $SU(m|n)$ supersymmetric HS model in 
(\ref{b6}) can be transformed to a spin chain. We consider a representation of
the permutation algebra (\ref{b5}), which acts on a spin state like $|\al_1 
\al_2 \dots \al_N \r$, with $\al_j \in \{1,2,\dots,m+n \}$, as 
\cite{basu2,basu3}
\beq \tilde{P}_{jk}^{(m|n)} |\al_1 \dots \al_j \dots \al_k \dots \al_N \r ~=~
e^{i \Phi (\al_j,\al_{j+1},\dots,\al_k)}~ |\al_1 \dots \al_k \dots \al_j \dots
\al_N \r. \label{b7} \eeq
Here 
$e^{i \Phi(\al_j,\al_{j+1},\dots,\al_k)}= 1$ if $\al_j,\al_k \in \{ 1,2,\dots,
m \}$, $e^{i \Phi(\al_j,\al_{j+1},\dots,\al_k)}= -1$ if $\al_j,\al_k \in 
\{ m+1,m+2,\dots,m+n \}$, and $e^{i \Phi(\al_j,\al_{j+1},\dots,\al_k)}= 
(-1)^{{\pi \sum_{p=j+1}^{k-1} \sum_{\tau=m+1}^{m+n} \de_{\al_p, \tau }}}$ 
if $\al_j \in \{ 1,2,\dots,m \}$ and $\al_k \in \{ m+1,m+2, \dots,m+n \}$ or 
vice versa. We will call $\al_i$ a `bosonic' spin if $\al_i \in \{ 1,2,\dots,
m \}$, and a `fermionic' spin if $\al_i \in \{ m+1,m+2,\dots,m+n \}$. From 
Eq. (\ref{b7}), it follows that the exchange of two bosonic (fermionic)
spins produces a phase factor of $ 1 (-1)$ irrespective of the nature of the
spins situated in between the $j$-th and $k$-th lattice sites. However, if
we exchange one bosonic spin with one fermionic spin, the phase factor is 
$(-1)^\rho$ where $\rho$ is the total number of fermionic spins situated in 
between the $j$-th and $k$-th lattice sites. The constraint in Eq. (\ref{b3})
implies that the Hilbert space associated with the $SU(m|n)$ HS Hamiltonian 
in (\ref{b6}) can be spanned through the following orthonormal basis vectors:
$C_{1 \al_1}^\da C_{2 \al_2}^\da \dots C_{N \al_N}^\da |0 \r $, where $|0 \r$
is the vacuum state and $\al_j \in \{ 1,2,\dots,m+n \}$. We define a 
one-to-one mapping between these basis vectors and those of the above 
mentioned spin chain as 
\beq |\al_1 \al_2 \dots \al_N \r ~\leftrightarrow ~C_{1 \al_1}^\da 
C_{2 \al_2}^\da \dots C_{N \al_N}^\da ~|0 \r . \label{b8} \eeq
Using the commutation (anticommutation) relations in (\ref{b2}), we can 
verify that 
\bea && \hat{P}_{jk}^{(m|n)} C_{1 \al_1}^\da \dots C_{j \al_j}^\da \dots
C_{k \al_k}^\da \dots C_{N \al_N}^\da |0 \r \nn \\
&& ~~~~~~~~~~~~~~~~~~~~~~~~~~ = e^{i\Phi(\al_j,\dots,\al_k)} C_{1 \al_1}^\da
\dots C_{j \al_k}^\da \dots C_{k \al_j}^\da \dots C_{N \al_N}^\da ~|0 \r ,
\label{b9} \eea
where $e^{i\Phi(\al_j,\dots,\al_k)}$ is the same phase factor which appeared 
in Eq. (\ref{b7}). A comparison of Eq. (\ref{b9}) with Eq. (\ref{b7}) 
through the mapping in (\ref{b8}) shows that the representation 
$\tilde{P}_{jk}^{(m|n)}$ is equivalent to the supersymmetric 
exchange operator $\hat{P}_{jk}^{(m|n)}$. Hence, if we define a spin 
chain Hamiltonian through $\tilde{P}_{jk}^{(m|n)}$ as
\beq H^{(m|n)}_{HS} ~=~ \frac{1}{2} ~\sum_{1 \le j<k \le N} ~
\frac{1+\tilde{P}_{jk}^{(m|n)}}{\sin^2(\xi_j-\xi_k)}, \label{b10} \eeq
it would be completely equivalent to the $SU(m|n)$ supersymmetric HS model in 
(\ref{b6}) \cite{basu2}. For the special case $n=0$, $\tilde{P}_{jk}^{(m|n)}$ 
reproduces the original spin exchange operator $P_{jk}$, and $H^{(m|n)}_{HS}$ 
in (\ref{b10}) reduces to the Hamiltonian of the $SU(m)$ HS spin chain in 
(\ref{a1}). We will henceforth study the $SU(m|n)$ supersymmetric HS 
model defined in (\ref{b10}) instead of its original form in (\ref{b6}).

Let us now discuss the partition function of the $SU(m|n)$ supersymmetric 
HS model, which has been derived by using the freezing technique 
\cite{basu4}. Consider a set of positive integers $k_1, k_2, \dots, k_r$, 
where $\sum_{i=1}^r k_i=N$, and $r$ is an integer which can take any value 
from $1$ to $N$. The vector ${\bf k} \equiv \{ k_1,\dots,k_r\}$
belongs to the set $\mathcal{P}_N$ of ordered partitions of $N$.
Associated with each ${\bf k}$, we attach a dimensionality given by
\beq d^{(m|n)}({\bf k}) ~=~ \prod_{i=1}^r d^{(m|n)}(k_i) \, , \label{b11} \eeq
where $d^{(m|n)}(k_i)$ is a function of $m$, $n$ and $k_i$. In the case of the
supersymmetric HS spin chain, for which both $m$ and $n$ are positive integers,
$d^{(m|n)}(k_i)$ is expressed as 
\beq d^{(m|n)}(k_i) =~ \sum_{j=0}^{{\rm min} (m,k_i)} {}^m C_j ~{}^{k_i-j+n-1}
C_{k_i-j} \, , \label{b12} \eeq
with ${}^p C_l=\frac{p!}{l\hskip .01 cm ! (p-l)!} $ for $l\leq p$ and ${}^p 
C_l=0 $ for $l> p$. In the case of the $SU(n)$ fermionic model, 
$d^{(0|n)}(k_i)$ is obtained by putting $m=0$ in Eq. (\ref{b12}),
\beq d^{(0|n)}(k_i) ~=~ {}^{k_i+n-1} C_{k_i}. \label{b13} \eeq 
The dimensionality of the $SU(m)$ bosonic case can also be obtained from Eq. 
(\ref{b11}) by taking \cite{fink}
\beq d^{(m|0)}(k_i)={}^m C_{k_i} \, . \label{b14} \eeq
We note that, while the dimensionality appearing in Eq. (\ref{b11}) can take 
a non-zero value for the bosonic case only if $k_i \le m $ for all $i$, it 
takes a non-zero value for any ${\bf k} \in \mathcal{P}_N$ for both the 
supersymmetric as well as the fermionic case.

Next we define the quantities $K_i = \sum_{j=1}^i k_j$ which denote the 
partial sums corresponding to the partition ${\bf k} \in \mathcal{P}_N$. 
The partition function of the $SU(m|n)$ HS spin chain, obtained through the 
freezing technique, is then given by \cite{basu4}
\beq Z^{(m|n)}_{HS}(q) ~=~ \prod_{l=1}^{N-1} \left(1- q^{\mathcal{E}(l)} 
\right)~ \sum _{\mathbf{k} \in ~\mathcal{P}_N} d^{(m|n)}({\bf k}) ~
\prod_{j=1}^{r-1} \frac{q^{\mathcal{E}(K_j) }}{(1-q^{\mathcal{E}(K_j) })}, 
\label{b15} \eeq
where $\mathcal{E}(l) = l(N-l)$ and $q = e^{-1/T}$; here $T$ is the 
temperature, and we have set the Boltzmann constant $k_B =1$. Note that the 
dimension of the summation variable $\mathbf{k}$ (i.e., $r$) in Eq. (\ref{b15})
takes all possible values within the range $1$ to $N$. Since the partial sums 
$K_1, K_2, \dots , K_r$ associated with ${\bf k}$ are natural numbers obeying 
$1 \le K_1< K_2< \dots<K_{r-1}<K_r = N$, one can define their complements as 
elements of the set: $\{1,2, \dots,N-1\}-\{K_1, K_2,\dots ,K_{r-1}\}$. Let 
$K_j$'s with $j\in \{ r+1,r+2, \dots ,N \} $ denote these conjugate partial 
sums. Hence one can rearrange the product $\prod_{l=1}^{N-1} 
(1-q^{\mathcal{E}(l)})$ into two terms as \cite{fink}
\beq \prod_{l=1}^{N-1} (1-q^{\mathcal{E}(l)}) ~=~ \prod_{j=1}^{r-1} 
(1-q^{\mathcal{E}(K_j) }) ~ \prod_{i=r+1}^{N}(1-q^{\mathcal{E}(K_i)}).
\label{b16} \eeq
By substituting this relation to Eq. (\ref{b15}), we get a simplified 
expression for the partition function of the $SU(m|n)$ HS model as
\beq Z^{(m|n)}_{HS}(q) ~=~\sum _{\mathbf{k} \in ~\mathcal{P}_N} d^{(m|n)} 
({\bf k})~ q^{\sum \limits^{r-1}_{j=1}\mathcal{E}(K_j) } ~\prod_{i=r+1}^{N}
(1-q^{\mathcal{E}(K_i) }). \label{b17} \eeq

Even though the partition function given in Eq. (\ref{b17}) is useful for 
studying various global properties of the spectrum like the level density 
distribution \cite{basu4}, it is not very suitable for analyzing the 
degeneracy of energy levels associated with the super-Yangian symmetry of the
$SU(m|n)$ HS model. However, it has been found recently that the partition 
function appearing in Eq. (\ref{b17}) can also be expressed as \cite{basu5}
\beq Z^{(m|n)}_{HS}(q) ~=~\sum _{\mathbf{k} \in ~\mathcal{P}_N} ~ 
q^{\sum \limits^{r-1}_{j=1}\mathcal{E}(K_j) } \, S_{\l k_1, k_2,\dots, k_r 
\r}(x,y) \rvert_{x=1,y=1} \, , \label{b18} \eeq
where $x \equiv \{x_1,x_2,\dots,x_m \}$, $y \equiv \{y_1,y_2,\dots,y_n \}$,
$\l k_1,k_2,\dots,k_r \r$ denotes a `border strip' which is drawn in Fig. 1,
and $S_{\l k_1, k_2, \dots, k_r \r}(x,y)$ is the Schur polynomial corresponding
to such a border strip.

These border strips represent a class of irreducible representations of the 
$Y(gl_{(m|n)})$ Yangian algebra, and they span the Fock space of Yangian 
invariant spin systems. These border strips can equivalently be described by 
motifs, which for an $N$-site spin chain is given by a sequence of $N-1$ number
of 0's and 1's, $\de = (\de_1, \de_2, \dots , \de_{N-1})$ with $\de_j \in \{ 
0, 1\}$. There exists a one-to-one map from a border strip to a motif as 
\beq \l k_1, k_2, \dots,k_r \r \, ~\Longrightarrow ~\de = (\,\underbrace{1,
\dots,1}_{k_1-1}\, , 0, \underbrace{1,\dots,1}_{k_2-1},0, \dots \dots,0,
\underbrace{1,\dots,1}_{k_r-1}\,) \, . \label{b19} \eeq
Thus the elements of this motif $\de$ satisfy the following rule: $\de_j = 0$
if $j$ coincides with one of the partial sums $K_i$, and $\de_j = 1$ otherwise.
The dimensionality of the irreducible representation associated with a border 
strip or motif is obtained by setting $x=1,~ y=1$ in the 
corresponding Schur polynomial $S_{\l k_1, k_2, \dots, k_r \r}(x,y)$.

There exist several alternative expressions for the Schur polynomial in the 
literature. For $x=1,~ y=1$, one such expression for the Schur polynomial 
\cite{basu5} is given by
\beq S_{ \l k_1,k_2,\dots,k_r \r}(x,y)\rvert_{x=1,y=1}~=~ \sum_{\mathbf{l} 
\in ~\mathcal{P}_r} (-1)^{r-s}\prod_{i=1}^s d^{(m|n)}\big( \sum_{j=1}^{\ell_i}
k_{\ell_1+\ell_2+\dots +\ell_{i-1}+j}\big) \, , \label{b20} \eeq
where the summation variable $\mathbf{l} \equiv \{\ell_1, \ell_2, \dots , 
\ell_s \}$ belongs to the set $\mathcal{P}_r$ of ordered partitions of $r$
(thus $s$ is an integer which runs from $1$ to $r$), and we assume that 
$l_0=0$. As an illustration, let us consider the Schur polynomial
$S_{ \l k_1,k_2,k_3 \r}(x,y)$, for which $r=3$ and $\mathcal{P}_3$ is given 
by $\{ \, \{ 3 \} , \{ 2, 1 \} , \{ 1,2 \}, \{ 1,1,1 \}\, \}.$ For 
$x=1,~ y=1$, Eq. (\ref{b20}) gives the value of this Schur polynomial to be
\bea && S_{ \l k_1,k_2, k_3 \r}(x,y)\rvert_{x=1,y=1} ~ = ~ 
d^{(m|n)} (k_1+k_2+k_3) \, - \, d^{(m|n)} (k_1+k_2)\, d^{(m|n)} (k_3) \nn \\
&& \hskip 2.5 cm - \, d^{(m|n)} (k_1) \, d^{(m|n)} (k_2+k_3) \,
+ \, d^{(m|n)} (k_1) \, d^{(m|n)} (k_2)\, d^{(m|n)} (k_3) \, . \nn \eea
It may be noted that, by using another combinatorial expression \cite{basu5}
for the Schur polynomial $S_{ \l k_1,k_2,\dots,k_r \r}(x,y)$, we find its 
value for $x=1,~ y=1$ to be
\beq S_{ \l k_1,k_2,\dots,k_r \r}(x,y) \rvert_{x=1,y=1} ~=~ 
N_{\l k_1,k_2,\dots,k_r \r } \, , \label{b21} \eeq
where $N_{\l k_1,k_2,\dots,k_r \r }$ denotes the number of all possible 
allowed tableaux corresponding to the border strip $\l k_1,k_2,\dots,k_r \r$.
An allowed tableau is obtained by filling the numbers $1, 2, \dots, m+n$ in a 
given border strip $\l k_1,k_2,\dots,k_r \r$ following the rules:
\begin{itemize}
\item The entries in each row are increasing, allowing the repetition of 
elements of the set $\{ 1,2, \dots , m\}$, but not permitting the repetition 
of elements of the set $\{ m+1, m+2, \dots , m+n \}$,

\item The entries in each column are increasing, allowing the repetition of
elements of the set $\{ m+1, m+2, \dots , m+n \}$, but not permitting the 
repetition of elements of the set $\{ 1,2, \dots , m\}$.
\end{itemize}
For example, in the case of the $SU(2|1)$ spin chain, it is possible to 
construct the following tableaux corresponding to the border strip $\l 2,1\r$:

\vskip .6 true cm
\begin{center}
\begin{tabular}{rrrrrrrr}
\setlength{\unitlength}{1pt}
\begin{picture}(40,50)(0,-50)
\put(0,-20){\line(1,0){20}}
\put(0,-10){\line(1,0){20}}
\put(10,0){\line(1,0){10}}
\put(20,0){\line(0,-1){20}}
\put(10,0){\line(0,-1){20}}
\put(0,-10){\line(0,-1){10}}
\put(10,-10){{\makebox(10,10){$1$}}}
\put(10,-20){{\makebox(10,10){$2$}}}
\put(0,-20){{\makebox(10,10){$1$}}}
\end{picture} &
\begin{picture}(40,50)(0,-50)
\put(0,-20){\line(1,0){20}}
\put(0,-10){\line(1,0){20}}
\put(10,0){\line(1,0){10}}
\put(20,0){\line(0,-1){20}}
\put(10,0){\line(0,-1){20}}
\put(0,-10){\line(0,-1){10}}
\put(10,-10){{\makebox(10,10){$1$}}}
\put(10,-20){{\makebox(10,10){$2$}}}
\put(0,-20){{\makebox(10,10){$2$}}}
\end{picture} &
\begin{picture}(40,50)(0,-50)
\put(0,-20){\line(1,0){20}}
\put(0,-10){\line(1,0){20}}
\put(10,0){\line(1,0){10}}
\put(20,0){\line(0,-1){20}}
\put(10,0){\line(0,-1){20}}
\put(0,-10){\line(0,-1){10}}
\put(10,-10){{\makebox(10,10){$1$}}}
\put(10,-20){{\makebox(10,10){$3$}}}
\put(0,-20){{\makebox(10,10){$1$}}}
\end{picture} &
\begin{picture}(40,50)(0,-50)
\put(0,-20){\line(1,0){20}}
\put(0,-10){\line(1,0){20}}
\put(10,0){\line(1,0){10}}
\put(20,0){\line(0,-1){20}}
\put(10,0){\line(0,-1){20}}
\put(0,-10){\line(0,-1){10}}
\put(10,-10){{\makebox(10,10){$1$}}}
\put(10,-20){{\makebox(10,10){$3$}}}
\put(0,-20){{\makebox(10,10){$2$}}}
\end{picture} &
\begin{picture}(40,50)(0,-50)
\put(0,-20){\line(1,0){20}}
\put(0,-10){\line(1,0){20}}
\put(10,0){\line(1,0){10}}
\put(20,0){\line(0,-1){20}}
\put(10,0){\line(0,-1){20}}
\put(0,-10){\line(0,-1){10}}
\put(10,-10){{\makebox(10,10){$2$}}}
\put(10,-20){{\makebox(10,10){$3$}}}
\put(0,-20){{\makebox(10,10){$1$}}}
\end{picture} &
\begin{picture}(40,50)(0,-50)
\put(0,-20){\line(1,0){20}}
\put(0,-10){\line(1,0){20}}
\put(10,0){\line(1,0){10}}
\put(20,0){\line(0,-1){20}}
\put(10,0){\line(0,-1){20}}
\put(0,-10){\line(0,-1){10}}
\put(10,-10){{\makebox(10,10){$2$}}}
\put(10,-20){{\makebox(10,10){$3$}}}
\put(0,-20){{\makebox(10,10){$2$}}}
\end{picture} &
\begin{picture}(40,50)(0,-50)
\put(0,-20){\line(1,0){20}}
\put(0,-10){\line(1,0){20}}
\put(10,0){\line(1,0){10}}
\put(20,0){\line(0,-1){20}}
\put(10,0){\line(0,-1){20}}
\put(0,-10){\line(0,-1){10}}
\put(10,-10){{\makebox(10,10){$3$}}}
\put(10,-20){{\makebox(10,10){$3$}}}
\put(0,-20){{\makebox(10,10){$1$}}}
\end{picture} &
\begin{picture}(40,50)(0,-50)
\put(0,-20){\line(1,0){20}}
\put(0,-10){\line(1,0){20}}
\put(10,0){\line(1,0){10}}
\put(20,0){\line(0,-1){20}}
\put(10,0){\line(0,-1){20}}
\put(0,-10){\line(0,-1){10}}
\put(10,-10){{\makebox(10,10){$3$}}}
\put(10,-20){{\makebox(10,10){$3$}}}
\put(0,-20){{\makebox(10,10){$2$}}}
\end{picture}
\end{tabular}
\end{center}
\vskip -1cm 
which gives $S_{ \l 2,1\r}(x,y)\rvert_{x=1,y=1}=8$. In the following, we shall
use both of the expressions in Eqs. (\ref{b20}) and (\ref{b21}) according to 
our convenience. 

Now we are in a position to find the energy levels and their degeneracies in 
the case of the $SU(m|n)$ HS spin chain. The energy level associated with a
border strip $\l k_1,k_2,\dots,k_r \r$ or corresponding motif $\de$ is obtained
from the power of $q$ appearing on the right hand side of Eq. (\ref{b18}):
\beq E (\de) ~=~ \sum_{i=1}^{r-1} ~\mathcal{E}(K_i) ~=~ \frac{N(N^2-1)}{6} ~+~
\sum_{j=1}^{N-1} ~\de_j ~j(j-N)\, . \label{b22} \eeq
{}From Eq. (\ref{b18}), it also follows that the degeneracy of the energy 
level $E(\de)$ associated with motif $\de$ is given by setting $x=1,~ y=1$ in 
the Schur polynomial ${S}_{ \l k_1,k_2,\dots,k_r \r}(x,y)$. We have already 
presented some expressions for this limit of the Schur polynomial in Eqs.
(\ref{b20}) and (\ref{b21}). It is interesting to observe that setting $x=1,~
y=1$ in the Schur polynomial $S_{ \l k_1,k_2,\dots,k_r \r}(x,y)$ also gives 
the degeneracy of the energy level associated with motif $\de$ in the 
case of the $SU(m|n)$ supersymmetric Polychronakos spin chain \cite {hika2}.

Let us suppose that there exists a unique motif $\de$ which minimizes $E(\de)$
and, therefore, represents the ground state of the $SU(m|n)$ HS spin chain. It
may be noted that, if $Z^{(m|n)}_{HS}(q)$ in Eq. (\ref{b17}) is expressed in a
polynomial form, then the term with the lowest power of $q$ is generated only 
from the partition ${\bf k}$ corresponding to the motif $\de$, and the 
coefficient of this power of $q$ is given by $d^{(m|n)}({\bf k})$. 
Consequently, the degeneracy of the motif $\de$ representing the ground 
state is obtained as 
\beq S_{ \l k_1,k_2,\dots,k_r \r}(x,y)\rvert_{x=1,y=1}~=~
d^{(m|n)}({\bf k}) ~=~ \prod_{i=1}^r d^{(m|n)}(k_i) \, . \label{b23} \eeq
This relation implies that only the vector $\mathbf{l} = \{1, 1, \dots , 1\}$
contributes to the right hand side of Eq. (\ref{b20}) for the case of the
ground state. If there exist more than one motif corresponding to the minimum 
energy of the system, then the degeneracy of the ground state is obtained by 
summing over all the $d^{(m|n)}({\bf k})$ associated with such motifs.

An interesting aspect of the HS models is a duality between the $SU(m|n)$ and 
$SU(n|m)$ spin chains of the form \cite{basu5}
\beq {\cal U} ~\tilde{P}_{jk}^{(m|n)} ~{\cal U}^\da = - \tilde{P}_{jk}^{(n|m)},
\label{b24} \eeq
and
\beq {\cal U} ~H^{(m|n)}_{HS} ~{\cal U}^\da = \frac{N(N^2-1)}{6} ~-~
H^{(n|m)}_{HS}, \label{b25} \eeq
where $\cal U$ is a unitary operator. It should be observed that, 
one can obtain the relation $\tilde{P}_{jk}^{(m|0)} = -\tilde{P}_{jk}^{(0|m)}$
by using Eq. (\ref{b7}). Hence, for the particular case $n=0$, 
$\cal U$ in Eqs. (\ref{b24}) and (\ref{b25}) acts like an unit operator 
between the Fock spaces of the $SU(m|0)$ bosonic and $SU(0|m)$ fermionic spin 
chains. [Note that the $SU(2)$ bosonic and fermionic spin chains correspond to
antiferromagnetic and ferromagnetic spin-1/2 chains respectively. This is 
because $\tilde{P}_{jk}^{(2|0)} = - \tilde{P}_{jk}^{(0|2)} = 2 {\vec S}_j 
\cdot {\vec S}_k + 1/2$, where ${\vec S}_j$ denotes a spin-1/2 operator at 
site $j$]. In general, Eq. (\ref{b25}) implies that if $| \psi_i \r$ is an 
eigenfunction of $H^{(m|n)}_{HS}$ with eigenvalue $E_i$, then ${\cal U} | 
\psi_i \r$ is an eigenfunction of $H^{(n|m)}_{HS}$ with eigenvalue $E_i'$, 
where 
\beq E_i'=\frac{N(N^2-1)}{6} ~-~E_i \, . \label{b26} \eeq
Furthermore, the degeneracy of the eigenvalue $E_i$ in the spectrum of 
the $SU(m|n)$ spin chain coincides with that of the eigenvalue $E_i'$ in the 
spectrum of the $SU(n|m)$ spin chain. Consequently, the partition functions of
the $SU(m|n)$ and $SU(n|m)$ spin chains satisfy the duality relation
\beq Z^{(m|n)}_{HS} (q) ~=~ q^{N(N^2-1)/6} ~Z^{(n|m)}_{HS} (q^{-1}) \, .
\label{b27} \eeq
For the special case of a `self dual' model with $m=n$, Eq. (\ref{b26}) 
implies that the spectrum would be symmetric under reflection around the
energy value $N(N^2-1)/12$. 

Let us now try to find out how the unitary operator ${\cal U}$ connects the 
motif representations appearing in the Fock spaces of the $SU(m|n)$ and 
$SU(n|m)$ HS spin chains. For this purpose, we need to define the conjugate 
of a border strip or corresponding motif. While the border strip
$ \l k_1,k_2,\dots,k_r \r $ has $r$ elements, its conjugate has 
$N-r+1$ elements which lead to a partition of $N$ \cite{basu5}. Thus
one can write this conjugate border strip as $\l k_1',k_2',\dots, k_{N-r+1}'
\r $, where $\{ k_1',k_2',\dots,k_{N-r+1}' \} \in \mathcal{P}_N$, and denote 
the partial sums corresponding to this conjugate border strip as $K_i'$ with 
$i\in \{ 1,2, \dots ,N-r+1\}$. The first $N-r$ such partial sums 
form a set, which is related to the complementary partial sums associated with
the original border strip $\l k_1,k_2,\dots,k_r \r$ as \cite{basu5}
\beq \{ K_1',K_2', \dots , K_{N-r}' \} = \{ N-K_{r+1}, N-K_{r+2}, \dots , 
N-K_N \} \, . \label{b28} \eeq
{}From the above equation, it follows that the conjugate of a motif can be 
obtained from the original motif by replacing $0$'s with $1$'s (and vice versa)
and rewriting all binary digits in the opposite order. For example, the 
conjugate of the motif $(10110)$ is obtained as $(10110) \to (01001) \to 
(10010)$. With the help of Eq. (\ref{b22}), we find that the eigenvalue of the
Hamiltonian $H^{(n|m)}_{HS}$ for the conjugate border strip $\l k_1',k_2',
\dots,k_{N-r+1}' \r $ is given by $\sum_{i=1}^{N-r} \mathcal{E} (K_i')$. Using
Eq. (\ref{b28}) along with the relations $\mathcal{E} (N-K_i)=\mathcal{E}(K_i)$
and $\sum_{i=1}^{N} \mathcal{E}(K_i) = N(N^2-1)/6$, we obtain 
\beq \sum_{i=1}^{N-r} \mathcal{E}(K_i')= \frac{N(N^2-1)}{6} - 
\sum_{i=1}^{r-1} \mathcal{E}(K_i) \, . \label{b29} \eeq
Comparing this equation with Eq. (\ref{b26}), we find that the unitary operator
${\cal U}$ maps the border strip $\l k_1,k_2,\dots,k_r \r$ appearing in the 
Fock space of the $SU(m|n)$ spin chain to the conjugate border strip $\l k_1',
k_2', \dots,k_{N-r+1}' \r$ appearing in the Fock space of the $SU(n|m)$ spin
chain. Furthermore, it is known that the Schur polynomials associated with
a border strip and its conjugate border strip satisfy a duality relation like 
\cite{basu5} 
\beq S_{ \l k_1,k_2,\dots,k_r \r}(x,y) ~=~ S_{\l k_1',k_2',\dots,k_{N-r+1}' \r}
(y,x) \, . \label{b30} \eeq
For $x=1, ~y=1$, this equation implies that the number of degenerate 
eigenfunctions of the $SU(m|n)$ HS spin chain associated with the motif $\l 
k_1, k_2,\dots,k_r \r$ coincides with that of the $SU(n|m)$ HS spin chain 
associated with the conjugate motif $\l k_1',k_2',\dots,k_{N-r+1}' \r$. This 
result is also consistent with our observation that a motif and its conjugate 
motif are related to each other through the unitary operator ${\cal U}$. 

It may be noted that the expression in (\ref{b22}) for the energy levels of the
$SU(m|n)$ HS spin chain does not explicitly depend on the values of $m$ and 
$n$, and apparently coincides with that of the $SU(m|0)$ bosonic case 
\cite{hald3}. However, as we shall see shortly, the degeneracy corresponding 
to some of these energy levels vanishes in the pure bosonic case
or pure fermionic case. As a result, the spectrum of a supersymmetric spin 
chain admits many more energy levels in comparison with the spectrum of a 
bosonic or fermionic spin chain with the same number of lattice sites. For 
the case of the $SU(m|0)$ bosonic spin chain, let us assume that the value of 
some $k_i$ in the border strip $\l k_1,k_2,\dots,k_r \r$ exceeds $m$.
Since it is not possible to construct a tableau corresponding to this border 
strip without the repetition of any number (from the set $\{1,2, \dots ,m\}$)
in the column which has the length $k_i$, we get $S_{ \l k_1,k_2,\dots,k_r 
\r}(x,y)\rvert_{x=1,y=1} =0$ by using Eq. (\ref{b21}). The same conclusion 
can also be drawn from Eq. (\ref{b20}) by observing that $d^{(m|0)}(l)=0$ for 
$l >m$. Consequently, by using Eq. (\ref{b19}), one finds a selection rule 
for the $SU(m|0)$ bosonic HS spin chain, which prohibits the occurrence of 
$m$ or more consecutive 1's in a motif. Note that if a motif contains a
sequence of $m$ or more consecutive 1's, then its conjugate 
motif would contain a sequence of $m$ or more consecutive $0$'s.
So, by using the duality relation in (\ref{b30}), we find a complementary 
selection rule for the $SU(0|m)$ fermionic HS spin chain, which prohibits the 
occurrence of $m$ or more consecutive $0$'s in a motif. In the case of the
$SU(m|n)$ supersymmetric spin chain, however, we can construct at least one 
tableau corresponding to the border strip $\l k_1,k_2,\dots,k_r \r$ for 
arbitrary values of $k_i$. The form of such a tableau has been shown in Fig. 2.

Consequently, by using Eq. (\ref{b21}), we find that 
${S}_{ \l k_1,k_2,\dots,k_r \r}(x,y)\rvert_{x=1,y=1} $ has a nonzero 
value for an arbitrary border strip $\l k_1,k_2,\dots,k_r \r$. 
Thus, the selection rules occurring in the bosonic and fermionic case are
lifted for the case of the supersymmetric HS spin chain; this was observed by 
Haldane on the basis of numerical calculations \cite{hald2}. 

Due to the absence of any selection rule, we can easily evaluate the maximum 
and minimum energy eigenvalues for the case of a supersymmetric HS spin chain.
{}From the expression for $E(\delta)$ in Eq. (\ref{b22}) it is evident that,
in the case of the $SU(m|n)$ supersymmetric as well as the $SU(0|n)$ fermionic
spin chain, the motif $\delta \equiv (1,1,\dots,1)$ corresponding to the 
border strip $\l N\r$ gives the minimum energy
of the system as $E_{min}= \frac{N(N^2-1)}{6} + \sum_{j=1}^{N-1} j(j-N)= 0$.
Due to the selection rule which prohibits the occurrence of $m$ or more 
consecutive 1's in a motif, the $SU(m|0)$ bosonic model has a nonzero ground 
state energy which will be discussed in Sec. 4. On the other hand, 
the motif $\delta \equiv (0,0,\dots,0)$ corresponding to the border strip 
$\l1,1,\dots ,1\r$ gives the maximum energy $E_{max}= \frac{N(N^2-1)}{6}$ for 
the $SU(m|n)$ supersymmetric as well as the $SU(m|0)$ bosonic spin chain. Due
to the presence of the selection rule which prohibits the occurrence of $m$ or
more consecutive 0's in a motif, the maximum energy of the $SU(0|m)$ fermionic
spin chain is lower than $N(N^2-1)/{6}$. The maximum energy of the $SU(0|m)$ 
fermionic spin chain can be obtained from the ground state energy 
of the $SU(m|0)$ bosonic spin chain by using the relation in (\ref{b26}). 

\no \section{Definition of momentum for the $SU(m|n)$ HS spin chain}
\renewcommand{\theequation}{3.{\arabic{equation}}}
\setcounter{equation}{0}

In order to study the energy-momentum dispersion relation, we need to find 
the momentum eigenvalues corresponding to the energy eigenstates of the 
$SU(m|n)$ HS spin chain. In this section, we will study how the momentum 
operator can be defined for the $SU(m|n)$ HS spin chain, and we will find 
the eigenvalues of this operator when it acts on the motif eigenstates. 

To begin with we consider the case of the $SU(m|0)$ bosonic 
spin chain, for which the momentum eigenvalues associated with the motif 
eigenstates are already known \cite{hald3}. The action of the translation 
operator $\mathcal{T}$ is defined on the spin space of this bosonic model as
\beq \mathcal{T} | \alpha_1, \alpha_2, \dots,\alpha_{N-1},\alpha_N \r = 
| \alpha_2, \alpha_3, \dots,\alpha_N, \alpha_1 \r . \label{c1} \eeq
$\mathcal{T}$ is an unitary operator which can be related to the 
momentum operator $\mathcal{P}$ as 
\beq \mathcal{T}= e^{i \mathcal{P}} \, . \label{c2} \eeq
It is well known that all degenerate energy eigenstates associated with the 
border strip $\l k_1,k_2,\dots,k_r \r$ or corresponding motif $\delta$ are 
also degenerate eigenstates of the translation operator $\mathcal{T}$ with
eigenvalue $e^{iP(\delta)}$, where $P(\delta)$ is given by \cite{hald3}
\beq P (\delta) \, = \, \big[ \,\pi (N-1) - \frac{2\pi}{N} \, 
\sum_{i=1}^{r-1} K_i \, \big] ~{\rm mod} ~2\pi \, = \, \frac{2\pi}{N} ~
\sum_{j=1}^{N-1} ~\delta_j ~j ~~~{\rm mod} ~2\pi. \label{c3} \eeq
Hence, due to the relation in (\ref{c2}), the momentum eigenvalue of all 
eigenstates associated with the motif $\delta$ is given by Eq. (\ref{c3})
for the case of a bosonic spin chain. 

Let us now define the translation and momentum operator for the general case 
of the $SU(m|n)$ HS spin chain. The translation operator acts on the creation 
(annihilation) operators associated with the Fock space of the $SU(m|n)$ spin
chain as 
\beq \mathcal{T} C_{i\alpha} \mathcal{T}^{\da} = C_{i'\alpha} \, ,~~~~~
\mathcal{T} C_{i\alpha}^\da \mathcal{T}^{\da} = 
C_{i'\alpha}^\da \, , \label{c4} \eeq
with $i' \equiv i+1$ (assuming $N+1 \equiv 1$ due to the circular 
configuration of the lattice sites), and on the vacuum state as 
$\mathcal{T}|0\r = |0\r$. Similar to the bosonic case, this translation 
operator is related to the momentum operator through Eq. (\ref{c2}). It is 
easy to check that the Hamiltonian in (\ref{b6}) of the $SU(m|n)$ HS spin 
chain commutes with the translation and momentum operator defined in the above 
mentioned way. Indeed, all conserved quantities \cite{hald2} of the $SU(m|n)$ 
HS model, which lead to the $Y(gl_{(m|n)})$ super-Yangian symmetry of this 
spin chain, also commute with these translation and momentum operators. This 
fact ensures that all degenerate energy eigenstates associated with the motif 
$\delta$ give the same momentum eigenvalue. With the help of Eqs. 
(\ref{c4}) and (\ref{b2}), we obtain 
\bea \mathcal {T} C_{1\alpha_1}^{\da}C_{2\alpha_2}^{\da}\dots 
C_{N-1 \, \alpha_{N-1}}^{\da} C_{N\alpha_N}^{\da}|0 \r 
&=& C_{2\alpha_1}^{\da}C_{3\alpha_2}^{\da}\dots
C_{N\alpha_{N-1}}^{\da}C_{1\alpha_N}^{\da}|0 \r \nn \\
&=& (-1)^{p(\alpha_N)\sum\limits_{j=1}^{N-1} p(\alpha_j) }
C_{1\alpha_N}^{\da} C_{2\alpha_1}^{\da} \dots 
C_{N\alpha_{N-1} }^{\da}|0 \r \, . \nn \\ \label{c5} \eea
Applying the mapping (\ref{b8}) to the above relation, we find the action of 
the translation operator $\mathcal{T}$ on the spin state $|\alpha_1, 
\alpha_2, \dots,\alpha_{N-1},\alpha_N \r$ as 
\beq \mathcal{T} | \alpha_1, \alpha_2, \dots,\alpha_{N-1},\alpha_N \r = 
(-1)^{p(\alpha_N)\sum\limits_{j=1}^{N-1} p(\alpha_j) }
| \alpha_2, \alpha_3, \dots,\alpha_N, \alpha_1 \r . \label{c6} \eeq
It may be noted that this general relation reduces to Eq. (\ref{c1}) in the 
particular case of the $SU(m|0)$ bosonic spin chain, for which $p(\alpha)=0$
for all possible values of $\alpha$. On the other hand, $p(\alpha)=1$ for all 
values of $\alpha$ in the case of the $SU(0|n)$ fermionic spin chain. Hence, 
in this case, Eq. (\ref{c6}) reduces to 
\beq \mathcal{T} | \alpha_1, \alpha_2, \dots,\alpha_{N-1},\alpha_N \r = 
e^{i\pi (N-1)} |\alpha_2,\alpha_3,\dots,\alpha_N,\alpha_1 \r . \label{c7} \eeq

Now we will find an expression for the momentum eigenvalue associated 
with a motif in the case of the $SU(0|n)$ fermionic spin chain, by using its 
duality relation with the $SU(n|0)$ bosonic spin chain. It has been already 
established that, any motif (say $\delta$) occurring in the Fock space of 
the $SU(0|n)$ fermionic spin chain can be related to its conjugate motif 
(say $\delta_c$) occurring in the Fock space of the $SU(n|0)$ bosonic spin 
chain through an unitary operator $\mathcal{U}$ which appears in Eqs. 
(\ref{b24}) and (\ref{b25}). More precisely, if $|\psi(\delta)\r$ is a state 
vector associated with the motif $\delta$, then there exists a state vector 
$|\psi(\delta_c)\r$ associated with the conjugate motif $\delta_c$ such that 
$|\psi(\delta)\r = \mathcal{U}|\psi(\delta_c)\r$. Since $\mathcal{U}$ acts 
like an unit operator in this case, we can express both $|\psi(\delta)\r$ and 
$|\psi(\delta_c)\r$ in exactly the same form through the corresponding basis 
vectors like $| \alpha_1, \alpha_2, \dots, \alpha_N \r \,$: 
\beq \sum_{\alpha_1, \alpha_2, \dots , \alpha_N} C_{\alpha_1, \alpha_2, 
\dots , \alpha_N} | \alpha_1, \alpha_2, \dots, \alpha_N \r \, , \label{c8} \eeq
where $C_{\alpha_1, \alpha_2, \dots , \alpha_N}$'s are some expansion 
coefficients. However, it should be kept in mind that while all $\alpha_i$ 
represent fermionic spins in the expression for $|\psi(\delta)\r$, they 
represent bosonic spins in the expression for $|\psi(\delta_c)\r$. Let us 
assume that $|\psi(\delta)\r$ and $|\psi(\delta_c)\r$ are eigenstates of the 
translation operator $\mathcal{T}$ with eigenvalues given by $e^{iP(\delta)}$ 
and $e^{iP(\delta_c)}$ respectively. Acting with $\mathcal{T}$ on the state 
vector appearing in Eq. (\ref{c8}), and using Eq. (\ref{c1}) or Eq. (\ref{c7})
in the case of bosons or fermions respectively, we find that 
\beq P(\delta) = \big[\pi (N-1) + P(\delta_c) \big] ~ {\rm mod} ~2\pi \, .
\label{c9} \eeq
With the help of Eqs. (\ref{c3}) and (\ref{b28}), one obtains the value of 
$P(\delta_c)$ as 
\beq P (\delta_c) =\, -\, \frac{2\pi}{N} ~\sum_{i=1}^{r-1} ~K_i ~~~{\rm mod} ~
2\pi ~ = \, \big[-\pi (N-1)+ \frac{2\pi}{N} \sum_{j=1}^{N-1} ~\delta_j ~j \, 
\big]~{\rm mod} ~2\pi \, , \label{c10} \eeq
where $K_i$'s are the partial sums associated with the border strip 
$\l k_1, k_2, \dots , k_r \r $, which has a one-to-one correspondence with 
the motif $\delta$. By inserting the above expression of $P (\delta_c)$
to Eq. (\ref{c9}), we find that the momentum eigenvalue of all eigenstates 
associated with the motif $\delta$ is again given by Eq. (\ref{c3}) for the 
case of the $SU(0|n)$ fermionic spin chain.

Our next aim is to find the momentum eigenvalues associated with the 
motifs of the $SU(m|n)$ supersymmetric spin chain, for which both $m$ and $n$ 
take nonzero values. For this purpose, we first consider the simplest case 
of the $SU(1|1)$ HS spin chain. It is well known that this spin chain can be 
mapped a model of non-interacting spinless fermions \cite{hald2,gohm}. Since 
the bosonic spin in the $SU(1|1)$ model can equivalently be described as a 
vacuum state for the fermionic spin, the corresponding exchange operator may 
be expressed as
\beq \tilde{P}_{jk}^{(1|1)} ~=~ 1 ~-~ C_j^\da C_j ~-~ C_k^\da C_k ~+~ C_j^\da
C_k ~+~ C_k^\da C_j, \label{c11} \eeq
where $C_j^\da (C_j)$ creates (annihilates) a spinless fermion at site $j$. 
The Hamiltonian in Eq. (\ref{b10}) can then be diagonalized in the form
\beq {H}^{(1|1)}_{HS} \, = \, \frac{N(N^2-1)}{6}-\sum_{u=0}^{N-1} 
~E_u ~{\tilde C}_u^\da {\tilde C}_u \, , \label{c12} \eeq
where $E_u =u (N-u)$, and ${\tilde C}_u$ is the Fourier transform of $C_j \,$:
\beq {\tilde C}_u = \frac{1}{\sqrt N} \sum_{j=1}^N e^{ 2i\pi uj/N } C_j \, .
\label{c13} \eeq
Let us consider a pure fermionic state of the following form:
\beq |\psi(u_1,u_2,\dots,u_r) \r \equiv \prod_{i=1}^r 
\tilde{C}^{\da}_{u_i} | 0 \r, \label{c14} \eeq
where $u_1<u_2< \dots < u_r$, and $u_i \in \{ 0,1,2,\dots,N-1 \}$. It is easy
to see that, this state is an eigenstate of the Hamiltonian in (\ref{c12}) 
with eigenvalue given by
\beq E(\{u_i\})=\frac{N(N^2-1)}{6}+\sum_{i=1}^r u_i(u_i-N). \label{c15} \eeq
Evidently, the set consisting of all states of the form given in (\ref{c14}) 
is a complete set of eigenstates for the Hamiltonian in (\ref{c12}).

Now we take an eigenstate of the form $|\psi(u_1,u_2,\dots,u_r) \r$, where
all $u_i$'s are positive integers. It may be noted that both 
$|\psi(u_1,u_2,\dots,u_r) \r$ and $|\psi(0,u_1,u_2,\dots,u_r) \r$ give rise 
to the same energy eigenvalue. So these eigenstates lead to a doubly 
degenerate energy level characterized by a set of fermionic quantum numbers 
like $\{ u_1,u_2,\dots,u_r \}$. It is natural to expect that such sets of 
quantum numbers would be connected in some way with the motifs of the 
$SU(1|1)$ HS spin chain. Indeed, by comparing two energy expressions given 
in Eqs. (\ref{b22}) and (\ref{c15}), we find that there exists a well defined 
mapping between the set $\{ u_1,u_2,\dots,u_r \}$ and the corresponding motif
$\delta$. The mapping rules are as follows:
\begin{itemize}
\item When $j \in \{u_1,u_2,\dots,u_r \}$, we have $\delta_j=1$.
\item When $j \notin \{u_1,u_2,\dots,u_r \}$, we have $\delta_j=0$.
\end{itemize}
To illustrate this mapping, we take a set of fermionic quantum numbers like 
$\{1,3,4\}$ for a spin chain with $N=6$. Due to the above mentioned rules, 
this set of quantum numbers is mapped to a motif of the form $(10110)$. 
Consequently, the degenerate eigenstates given by 
$\tilde{C}^{\da}_1\tilde{C}^{\da}_3\tilde{C}^{\da}_4|0\r$ 
and $\tilde{C}^{\da}_0 \tilde{C}^{\da}_1 \tilde{C}^{\da}_3
\tilde{C}^{\da}_4|0\r$ correspond to the motif $(10110)$. 

We have already defined the translation and momentum operators for a 
supersymmetric spin chain. As will be shown shortly, for the case of the
$SU(1|1)$ spin chain, these operators can be expressed in simple forms through
Fourier transformed modes like $\tilde{C}_u$ and $\tilde{C}^{\da}_u$.
By using Eqs. (\ref{c4}) and (\ref{c13}), we find the action of translation 
operator on $\tilde{C}_u$ and $\tilde{C}^{\da}_u$ as
\beq \mathcal{T} \tilde{C}_{u} \mathcal{T}^{\da} = e^{-i 2 \pi u/N} 
\tilde{C}_{u} \, ,~~~~~ \mathcal{T} \tilde{C}_{u}^{\da} 
\mathcal{T}^{\da} = e^{i 2\pi u /N} \tilde{C}_{u}^{\da} \,.
\label{c16} \eeq
With the help of the Baker-Hausdorff relation, it is easy to check that a 
translation operator of the form given in (\ref{c2}) satisfies Eq. (\ref{c16}) 
if the momentum operator $\mathcal{P}$ is given by 
\beq \mathcal{P}=\frac{2 \pi}{N} ~\sum_{u=0}^{N-1} ~u~\tilde{C}_u^\da
\tilde{C}_u \, . \label{c17} \eeq
Acting on the states $|\psi(u_1,u_2,\dots,u_r) \r$ and 
$|\psi(0,u_1,u_2,\dots,u_r) \r$, this momentum operator evidently 
generates the same eigenvalue given by 
\beq P(\{u_i\})=\frac{2\pi}{N}\sum_{i=1}^r u_i ~ ~\text{mod}~ 2\pi \, .
\label{c18} \eeq
By utilizing the mapping between fermionic quantum numbers $\{ u_1,u_2,\dots,
u_r \}$ and motif $\delta$, we can also express the momentum eigenvalue in 
Eq. (\ref{c18}) through the elements of motif $\delta$. Interestingly, we 
find that such an expression for the momentum eigenvalue for the 
motif $\delta$ is identical in form with the expression in (\ref{c3}), which 
was originally proposed for the case of $SU(m|0)$ bosonic spin chain. 

Finally, let us discuss how the momentum eigenvalue relations obtained for 
the simplest case of the $SU(1|1)$ supersymmetric spin chain can be useful in 
the context of the general $SU(m|n)$ supersymmetric spin chain. It should be 
observed that, for a fixed number of lattice sites, all possible motifs of 
the $SU(m|n)$ supersymmetric spin chain also occur in the case of the 
$SU(1|1)$ spin chain. However, while all motifs are doubly degenerate in the 
case of the $SU(1|1)$ spin chain, the number of degenerate eigenstates 
associated with a motif is much higher in general in the case of the $SU(m|n)$
supersymmetric spin chain. In fact, the doubly degenerate eigenstates 
associated with a motif $\delta$ of the $SU(1|1)$ spin chain form a subset of
the multiply degenerate eigenstates associated with the {\it same} motif 
$\delta$ of the $SU(m|n)$ spin chain. We have already found that, the momentum
eigenvalue of this subset of $SU(1|1)$ doublets is given by $P(\delta)$ in 
Eq. (\ref{c3}). Since all degenerate multiplets associated with the motif 
$\delta$ of the $SU(m|n)$ supersymmetric spin chain must yield the same 
momentum eigenvalue, it is also given by Eq. (\ref{c3}). Thus we find that 
$P(\delta)$ given in Eq. (\ref{c3}) represents a general expression for the 
momentum eigenvalue corresponding to the motif $\delta$, which is valid for all
possible cases like the bosonic, fermionic and supersymmetric HS spin chains. 

\no \section{Low energy excitations of the $SU(m|n)$ HS spin chain}
\renewcommand{\theequation}{4.{\arabic{equation}}}
\setcounter{equation}{0}

In this section, we will study the ground state and low energy excitations
of the $SU(m|n)$ HS spin chain for various values of $m$ and $n$. In
particular, we will be interested in the thermodynamic limit $N \to \infty$.
We will therefore rescale the Hamiltonian to take the form
\beq H^{(m|n)}_{HS} ~=~ \frac{\pi^2}{2N^2} ~\sum_{1 \le j<k \le N} ~
\frac{1+\tilde{P}_{jk}^{(m|n)}}{\sin^2 (\xi_j-\xi_k)}. \label{d1} \eeq
The pre-factor of $\pi^2 /(2N^2)$ in (\ref{d1}) ensures that the 
nearest-neighbor interaction is of the form $(1+\tilde{P}_{j,j+1}^{(m|n)})/2$,
and also that the ground state energy per unit length will remain finite as 
$N \to \infty$. (We have set the lattice spacing equal to 1). In that 
limit, Eq. (\ref{d1}) can be re-written as
\beq H^{(m|n)}_{HS} ~=~ \frac{1}{2} ~\sum_{j<k} ~\frac{1+\tilde{P}_{jk}^{
(m|n)}}{(j-k)^2} \label{d2} \eeq 
for $|j - k| \ll N$. The partition function corresponding to the rescaled 
Hamiltonian in (\ref{d1}) may be obtained from the expressions in (\ref{b17}) 
or (\ref{b18}) after replacing $q$ by $\tq$, where $\tq = e^{-\pi^2 /N^2T}$. 

Using the results of the previous section for the case of the rescaled 
$SU(m|n)$ spin chain, we can express its energy and momentum eigenvalues 
corresponding to the motif $\delta$ as some functions of the integers $K_i$,
namely,
\bea E &=& \frac{\pi^2}{N^2} ~\sum_{i=1}^{r-1} ~K_i (N - K_i) , \nn \\
{\rm and} ~~~P &=& \big[\, \pi (N-1) ~-~ \frac{2\pi}{N} ~
\sum_{i=1}^{r-1} ~K_i \, \big] ~~~{\rm mod} ~2\pi \, .\label{d3} \eea
For the Hamiltonian given in Eq. (\ref{d1}), the low energy modes are those 
for which $E$ is of order $1/N$, while the high energy modes are those for 
which $E$ is of order 1. Let us now consider three cases separately.

\vskip .4 true cm
\no {\bf Case I: $SU(m|0)$ bosonic spin chain, with $m \ge 2$}. 
\vskip .1 true cm

Although this bosonic spin chain has been discussed extensively in the 
literature, we will consider it briefly for the sake of completeness. 
Let us consider the simplest case when $N$ is a multiple of $m$. In this case,
the border strip $\l m,m, \dots , m \r$ minimizes the energy $E$ in Eq. 
(\ref{d3}) and represents the ground state of the system. For this border 
strip, we have $r=N/m$, $k_j = m$ for all $j \in \{ 1,2,\dots, r \}$, and $K_j
= jm$ for all $j \in \{ 1,2,\dots, r-1 \}$. Due to Eqs. (\ref{b23}) and
(\ref{b14}), the ground state is non-degenerate for this case.
Eq. (\ref{d3}) gives the ground state energy and momentum to be
\bea E_0 &=& \frac{\pi^2 m}{6N} ~\left[ \left( \frac{N}{m} \right)^2 ~-~ 1 
\right], \nn \\
{\rm and} ~~~P_0 &=& \pi N (1 ~-~ \frac{1}{m}) ~~~{\rm mod} ~2 \pi . 
\label{d4} \eea
In the thermodynamic limit, the ground state energy per unit length is given 
by $\pi^2 /(6m)$.

A band of excited states is obtained by taking $r=N/m + 1$ as follows. In 
addition to the $N/m-1$ values of $K_j = jm$ that are present in the ground 
state, we introduce an additional value of $K_i = u$, where $u$ is not a 
multiple of $m$. The excitation energy and momentum of such a state, called
$\Delta E$ and $\Delta P$ respectively, are given by the differences between 
the energy and momentum of this state and the ground state energy and momentum
given in Eq. (\ref{d4}). We find that $\Delta E = (\pi^2 / N^2) u(N-u)$, while
$\Delta P = - 2\pi u/N$ mod $2\pi$. In the thermodynamic limit $u/N \to 0$ or 
$(N-u)/N \to 0$, these correspond to excitations with low energy and low 
momentum. The velocity corresponding to these excitations is given by $v = 
|\Delta E/ \Delta P|$ which is equal to $\pi /2$.

It is interesting to observe that the excitations described above do {\it not}
have the lowest possible value of $\Delta E$, even for $u=1$ or $N-1$. Rather,
the excitations with the lowest energy are given by $r=N/m + 1$ and $K_j = 
(j-1)m+u$ for all $j \in \{ 1,2,\dots, r-1 \}$, where $u=1$ or $m-1$. These 
excitations have $\Delta E= (\pi^2 /N^2) N(1-1/m)$, $\Delta P = \pm 2 \pi /m$
mod $2\pi$, and a degeneracy of $m^2$. For $N \to \infty$, these excitations 
have less energy than the lowest energy excitations described in the previous 
paragraph, although their momentum is large, i.e., $2\pi /m$ instead of $2\pi
/N$. The momentum value of these excitations suggests that they may be related
to the algebraic long-range order which is exhibited by the ground state of 
this model; as discussed in Eq. (\ref{e1}) below, the two-point correlation 
oscillates with a wave number $2\pi /m$ and decays as a power of the distance.
Finally, the ratio of the energy of these high momentum excitations to the 
lowest energy of the low momentum excitations is $1 - 1/m$. 
We note that low energy states with momentum equal to $\pi$ are also known to 
exist in the spin-1/2 $XXZ$ chain with nearest-neighbor couplings \cite{suth2}.

\vskip .4 true cm
\no {\bf Case II: $SU(0|n)$ fermionic spin chain, with $n \ge 2$}.
\vskip .1 true cm

We have already seen in Sec. 2 that the border strip $\l N\r$ or 
corresponding motif $(11\dots 1)$ represents the ground state of this system.
The ground state energy is zero and momentum is given by $\pi (N-1)$. Eqs. 
(\ref{b23}) and (\ref{b13}) yield the ground state degeneracy as ${}^{N+n-1} 
C_N$; this goes as $N^{n-1} /(n-1)!$ in the thermodynamic limit.

The ground state and the vanishing of its energy can be understood as follows.
The simplest ground state is given by a state in which every site has a 
fermionic spin of the same type, say, $\al = 1$, using the notation given at 
the beginning of Sec. 2 and remembering that $m=0$. The arguments given after 
Eq. (\ref{b7}) imply that when $\tilde{P}_{jk}^{(0|n)}$ acts on such a state,
it gives $-1$ for all values of $j$ and $k$. Hence the state has zero 
eigenvalue for the Hamiltonian in Eq. (\ref{d1}). One can now see the form of
the general ground states. Consider a state in which the first $n_1$ sites 
have fermionic spins of type $\al = 1$, the next $n_2$ sites have spins of 
type $2$, and so on, with the condition that $\sum_{j=1}^n n_j = N$. One can 
then consider a superposition of states, all having the same amplitude, in 
which the same set of spins is distributed in all possible ways over all the 
lattice sites. This gives a ground state with zero energy since $1 + 
\tilde{P}_{jk}^{(0|n)}$ acting on such a superposition gives zero for all 
pairs $j$ and $k$. The number of ways of choosing $n$ ordered integers (some 
of which can be zero) which add up to $N$ is equal to ${}^{N+n-1} C_N$; this 
is the degeneracy of the ground state.

A band of excited states is obtained by taking border strips like $\l k_1 , 
k_2 \r$, with $k_1 = N-u$, $k_2 = u$, and $K_1 =N-u$, where $1 \le u \le N-1$.
The excitation energy and momentum of this state are given by $\Delta E = 
(\pi^2 /N^2) u(N-u)$ and $\Delta P = P- \pi (N-1) = 2\pi u/N$ respectively. 
The velocity is given by $v = |dE/dP| = \pi /2$ in the thermodynamic limit 
$u/N \to 0$ or $(N-u)/N \to 0$.

The wave function of these excited states can be visualized by considering a 
state $|j \r$, in which the site labeled $j$ is occupied by a fermionic spin 
of type, say, $\al = 2$, while all the other sites are occupied by fermionic 
spins of type $\al = 1$. We form a state with wave number $u$ by superposing 
such states,
\beq |u \r ~=~ \frac{1}{\sqrt N} ~\sum_{j=1}^N ~e^{i2\pi uj/N} ~|j \r, 
\label{d5} \eeq
where $u \in \{ 1, 2, \dots, N-1 \}$. We then find that this state is an 
eigenstate of the Hamiltonian in Eq. (\ref{b10}) with the eigenvalue
\beq E_u ~=~ \frac{\pi^2}{2N^2} ~\sum_{j=1}^{N-1} ~\frac{1 ~-~ \cos (2 \pi 
uj/N)}{\sin^2 (\pi j/N)} ~=~ \frac{\pi^2}{N^2} ~u ~(N - u), \label{d6} \eeq
where the last equality can be found in Ref. \cite{hald1}. Since the energy 
of the ground state is zero, the excitation energy of the state $|u\r $ 
is also given by $E_u$. In the limit $N \to \infty$, the excitation momentum 
$\Delta p= 2 \pi u/N$ becomes a continuous variable lying in 
the range $(0, 2 \pi)$, and we obtain the dispersion 
\beq \Delta E = \frac{\pi \Delta p}{2} ~-~ \frac{(\Delta p)^2}{4}. 
\label{d7} \eeq
The excitation energy $\Delta E$ goes to zero linearly as $\Delta p \to 0$ 
or $2 \pi$, with a slope given by the velocity $v = |dE/dp| = \pi /2$.

Note that the linear dispersions near $\Delta p=0$ and $2\pi$ in Eq. 
(\ref{d7}) are due to the $1/j^2$ interaction between pairs of sites 
separated by a distance $j$. If the interaction was short-ranged, the 
dispersion would be quadratic near $\Delta p=0$ and $2\pi$; this is known to
be the case for a ferromagnetic spin chain with nearest-neighbor interactions.

\vskip .4 true cm
\no {\bf Case III: $SU(m|n)$ spin chain, with $m, n \ge 1$}.
\vskip .1 true cm

Similar to case II, the border strip $\l N\r$ or corresponding motif $(11 \dots
1)$ represents the ground state of this system. The ground state energy is 
zero and momentum is given by $\pi (N-1)$. (Note that all the ground states 
of case II are also ground states of case III). However, due to the presence 
of bosons, the degeneracy of the ground state would be different from that of 
the $SU(0|n)$ fermionic case. By using Eqs. (\ref{b12}) and (\ref{b23}), and 
also assuming $N\ge m$, we find that the ground state degeneracy of the 
$SU(m|n)$ spin chain is given by 
\beq \sum_{j=0}^m ~{}^m C_j ~{}^{N+n-j-1} C_{N-j}, \label{d8} \eeq
which goes as $2^m N^{n-1} /(n-1)!$ in the thermodynamic limit. 

It is interesting to observe that, for the special case $n=1$, the ground 
state degeneracy given in Eq. (\ref{d8}) reduces to $2^m$, independent of $N$.
This can be understood as follows. The ground states consist of each of the 
$m$ types of bosonic spins either not appearing at all, or appearing in only 
one site of the chain; this gives rise to $2^m$ possibilities. Thus $p$ sites 
of the chain have bosonic spins, where $0 \le p \le m$, while the remaining 
$N-p$ sites are occupied by the fermionic spins.

A band of low energy excitations are obtained by taking border strips like 
$\l N-u , u \r$, where $1 \le u \le N-1$. Again, by following a method similar
to Case II which leads to the state vector $|u \r$ in Eq. (\ref{d5}), the wave
function of these excited states can be constructed explicitly. Consequently, 
the excitation energy and momentum of these states are obtained as 
$\Delta E=(\pi^2 /N^2) u(N-u)$ and $\Delta P = 2\pi u/N$ 
respectively. Thus we find that the low lying energy levels have the same 
motif structure and momentum for the $SU(m|n)$ supersymmetric as well as 
the $SU(0|n)$ fermionic case. In the thermodynamic limit, these low lying 
excitations have a linear dispersion, with the velocity being given by $\pi/2$.

For the $SU(m|1)$ spin chain, it is worth noting that the degeneracy of both 
the ground state and all the low lying excited states contains a factor of 
$2^m$. According to Eqs. (\ref{b12}) and (\ref{b20}), the degeneracy of all 
states for which at least one of the $k_i \ge m$ will contain a factor of 
$2^m$. The states whose degeneracy is not a multiple of $2^m$ are the ones 
in which all the $k_i < m$; these states necessarily have energies of order 
1, and are therefore high energy states.

It is interesting to consider what happens if the couplings of the 
$SU(m|n)$ spin chain are not of the inverse square form, i.e., not of the 
Haldane-Shastry type. Let us restrict our attention to the case $n \ge 1$
and take a rather general form of a $SU(m|n)$ symmetric Hamiltonian like 
\beq \mathcal{H}^{(m|n)} ~=~ \sum_{j<k} ~w (|k-j|) ~( 1 ~+~ 
\tilde{P}_{jk}^{(m|n)}), \label{d9} \eeq
where all the $w (j)$'s are arbitrary real positive numbers. We have already 
seen that the ground states of the HS spin chain Hamiltonian in Eq. (\ref{d1})
have zero energy for $n \ge 1$; hence they satisfy $\tilde{P}_{jk}^{(m|n)} = 
-1$ for all pairs of sites $j,k$. Clearly, such states will continue to be 
ground states of $\mathcal{H}^{(m|n)}$ in Eq. (\ref{d9}), since $w (j) > 0$ 
and the minimum eigenvalue of $\tilde{P}_{jk}^{(m|n)}$ is $-1$. Consequently,
we find that the ground states of the $SU(m|n)$ symmetric Hamiltonian 
$\mathcal{H}^{(m|n)}$ and the degeneracy of these states remain the same 
even if $w (j)$ is not of the inverse square form. This conclusion is not 
surprising because the degeneracy of the ground state of the $SU(m|n)$ 
supersymmetric HS model is governed by the border strip $\l N \r$, which 
coincides with the single column Young diagram $[1^N]$ associated with the
$SU(m|n)$ super Lie algebra. 

However, the energy-momentum relation for the low energy excitations of  
$\mathcal{H}^{(m|n)}$ will not be linear in general. Following arguments 
similar to Eqs. (\ref{d5}-\ref{d7}), we see that the energy of a state with 
momentum $p$ (measured with respect to a ground state) is given by 
\beq E_p ~=~ \sum_{j=1}^\infty ~w (j) ~[ 1 ~-~ cos (pj) ]. \label{d10} \eeq
Although $E_p$ will go to zero at $p = 0$ and $2\pi$, the dispersion $E_p$ 
versus $p$ near those two points is not linear in general. For instance, if
$w (j)$ decreases exponentially with $j$ as $j \to \infty$, the dispersion
is quadratic. As we will discuss in the next section, the low-energy 
excitations of a one-dimensional model cannot be described by a conformal 
field theory if the dispersion is not linear; hence the low-energy excitations
of the Hamiltonian $\mathcal{H}^{(m|n)}$ will not be governed by a conformal 
field theory unless $w(j)$ is chosen in some specific way like inverse square
interaction. 

\no \section{Conformal field theory description of low energy excitations}
\renewcommand{\theequation}{5.{\arabic{equation}}}
\setcounter{equation}{0}

Having discussed the low energy excitations of the $SU(m|n)$ HS spin chain for
different values of $m$ and $n$, we will now examine if these excitations can
be described by a conformal field theory (CFT) in the thermodynamic limit
\cite{itzy}. A CFT must have a finite number of ground states and a linear 
energy-momentum relation for the excitations. The first property implies that 
only the $SU(m|0)$ bosonic spin chain with $m \ge 2$ and the $SU(m|1)$ spin 
chain with $m \ge 1$ can possibly be governed by some CFTs. The central charges
for the $SU(m|0)$ and $SU(m|1)$ Polychronakos spin chains were calculated in 
Ref. \cite{hika2}, and were found to be $m-1$ and $m/2$ respectively; we should
point out that the convention for $(m|n)$ followed in \cite{hika2} is the 
reverse of the convention that we are following here.

Let us first consider the $SU(m|0)$ HS spin chain briefly. It is known 
\cite{hald3,scho,bouw} that the low energy excitations of this model are 
governed by the $SU(m)_1$ Wess-Zumino-Novikov-Witten (WZNW) model
\cite{witt,poly4,kniz}. In particular, the central charge of the $SU(m|0)$ 
HS spin chain is given by $c=m-1$. In fact, Ref. \cite{affl1} had already 
identified the $SU(m)_1$ WZNW model as providing a description of the 
low energy excitations of a class of $SU(m)$ symmetric spin chains. An 
important property of such a CFT is that the two-point equal-time 
correlation function in the ground state $|G \r$ goes as
\beq \sum_{\al,\be =1}^m ~\l G| S_{\al \be} (j_1) S_{\be \al} (j_2) |G \r ~
\sim ~ \frac{\cos [2\pi (j_1 - j_2)/m ]}{|j_1 - j_2|^{2-2/m}} \label{e1} \eeq
for $|j_1 - j_2| \to \infty$, where $S_{\al \be} (j) = C_{j\al}^\da C_{j\be}
- \de_{\al \be}/m$ \cite{affl1}. (Note that for the $SU(m|0)$ spin chain, 
the constraint in (\ref{b3}) implies that $\sum_{\al =1}^m \l G| S_{\al \al}
(j) |G \r= 0$ for all $j$). The period $m$ of the oscillations in (\ref{e1})
is consistent with the observation that the size of the system defined on
a lattice must be a multiple of $m$ in order to have a unique ground state. 

We now turn to the $SU(m|1)$ supersymmetric spin chain. As we saw earlier, 
the ground state degeneracy is $2^m$. In all the ground states, most of the 
sites are occupied by a fermionic spin, and only $p$ sites are occupied by 
bosonic spins, where $0 \le p \le m$. It is therefore convenient to perform 
a duality transformation given by Eq. (\ref{b24}) to obtain a 
$SU(1|m)$ spin chain governed by the Hamiltonian
\beq {\bar H}^{(1|m)} ~=~ \frac{\pi^2}{2N^2} ~\sum_{1 \le j<k \le N} ~
\frac{1-\tilde{P}_{jk}^{(1|m)}}{\sin^2 (\xi_j-\xi_k)}, \label{e2} \eeq
so that the ground states will have most of the sites occupied by a bosonic
spin. Further, since there is only one kind of bosonic spin, we can think
of it instead as a vacuum state for the fermions. We will now study the 
model defined in Eq. (\ref{e2}) for different values of $m \ge 1$.

We have already mentioned in Sec. 3 that the $SU(1|1)$ spin chain is 
equivalent to a model of non-interacting spinless fermions. This is because
for $m=1$, the exchange operator appearing in Eq. (\ref{e2}) can be written 
in the form
\beq \tilde{P}_{jk}^{(1|1)} ~=~ 1 ~-~ D_j^\da D_j ~-~ D_k^\da D_k ~+~ D_j^\da
D_k ~+~ D_k^\da D_j, \label{e3} \eeq
where $D_j^\da (D_j)$ creates (annihilates) a fermion at site $j$. The 
Hamiltonian in Eq. (\ref{e2}) can then be diagonalized; it takes the form
\bea {\bar H}^{(1|1)} &=& \sum_{u=0}^{N-1} ~E_u ~{\tilde D}_u^\da 
{\tilde D}_u, \nn \\
{\rm where} ~~~E_u &=& \frac{\pi^2}{N^2} ~u ~ (N-u), \label{e4} \eea
and ${\tilde D}_u$ is the Fourier transform
of $D_j$. Note that the mode with $u=0$ has zero energy, while all the other 
modes have positive energy. The ground state corresponds to all the positive 
energy states being empty. The zero energy state can be either filled or 
empty; this gives rise to a two-fold degeneracy of the ground state. In the 
thermodynamic limit, we define an excitation momentum $\Delta p= 2\pi u/N$ as 
usual. The low energy excitations have a dispersion which is linear near 
$\Delta p=0$ and $2\pi$. Near these two points, the momentum, which is defined
mod $2\pi$, is restricted to positive and negative values respectively, and 
the dispersions are given by $dE/dp = \pm v$ respectively. 

It is interesting to note that the fermionic operators appearing in Eqs. 
(\ref{e3}) and (\ref{c11}) are related by a particle-hole symmetry which 
implements the duality given in Eq. (\ref{b24}) for $m=n=1$. Consider the 
unitary operator $U = \exp [i(\pi /2) \sum_j (C_j D_j + D_j^\da C_j^\da)]$, 
where the $C_j$'s and $D_j$'s are independent fermion operators which 
anticommute with each other. We find that $U C_j U^\da = i 
D_j^\da$ and $U C_j^\da U^\da = - i D_j$. We can then verify that an 
unitary transformation by $U$ relates the exchange operators in Eqs. 
(\ref{e3}) and (\ref{c11}) in such a way as to satisfy Eq. (\ref{b24}). 

Due to the exact equivalence of the $SU(1|1)$ HS spin chain to a system of 
fermions given in Eq. (\ref{e4}) for any value of $N$, the partition function
is exactly given by $Z^{(1|1)}_{HS}(\tq) = Z_1$, where
\beq Z_1 ~=~ \prod_{u=0}^{N-1} ~( 1 ~+~ \tq^{ \, u(N-u)} ), \label{e5} \eeq
and $\tq = e^{-\pi^2 /(N^2 T)}$. A different proof of this identity is given in
Appendix A. Let us now consider the thermodynamic limit of the model. In this
limit, we can define an excitation momentum $\Delta p=2\pi u/N$ at $u/N \to 
0$ or $\Delta p=- 2\pi (N-u)/N$ at $(N-u)/N \to 0$. Taking these two kinds of 
low energy modes together, the partition function at low temperatures (i.e., 
$T \ll 1$ in the units we are using) is given by
\beq \ln Z_1 ~=~ 2 \int_0^\infty \frac{dp}{2\pi /N} ~\ln (1 ~+~ e^{-vp/T}),
\label{e6} \eeq
where $v=\pi /2$ is the velocity. The energy per unit length is given by
$(T^2/N) \partial \ln Z_1 /\partial T$, and we find that this is equal to $\pi
T^2 /(12v)$. The specific heat/length is obtained by differentiating this with
respect to $T$, and therefore equals $\pi T/(6v)$. One can also evaluate the 
average number of fermions per unit length. We introduce a chemical potential 
$\mu$ in Eq. (\ref{e6}) by replacing $e^{-vp/T}$ by $e^{-(vp-\mu) /T}$; the 
number of fermions per unit length is then given by $(T/N) (\partial \ln Z_1 /
\partial \mu)_{\mu=0}$, and it turns out to be equal to $T \ln 2 /(\pi v)$.

For a CFT with central charge $c$, the specific heat/length equals $\pi 
cT/(3v)$ \cite{blot,affl2}. The central charge of the $SU(1|1)$ HS spin chain
is therefore given by $c=1/2$. Note that this is half the central charge $c=1$
of a massless Dirac fermion; the latter has both positive and negative energy 
modes with two different linear dispersions $\Delta E = \pm v \Delta p$, where
$\Delta p$ can go from $-\infty$ to $\infty$ in both cases. The low energy 
modes of the $SU(1|1)$ HS spin chain only have positive energies, and can 
therefore be thought of as the modes of half of a Dirac fermion.

We will now consider the system defined in Eq. (\ref{e2}) for $m \ge 2$. We 
will argue that in the limit $N \to \infty$ and temperatures $T \ll 1$, this 
system is equivalent to a model of $m$ species of non-interacting fermions. 
This can be physically understood as follows. By using Eq. (\ref{b7}) for the 
special case $m=1$, and interpreting the only one kind of bosonic spin 
occurring in this case as a hole for the fermions, the exchange operator in 
Eq. (\ref{e2}) can be written as
\bea \tilde{P}_{jk}^{(1|m)} &=& 1 \, + \, \sum_{\al=1}^m ~\big[\, - \, 
D_{j\al}^\da D_{j\al} \, - \, D_{k\al}^\da D_{k\al} \, +\, 
D_{j\al}^\da D_{k\al} ~+~ D_{k\al}^\da D_{j\al} \, \big] \nn \\ 
& & + \sum_{1 \le \al \ne \be \le m} ~\big[ \, D_{j\al}^\da D_{k\be}^\da 
D_{j\be} D_{k\al}\, - \, D_{j\al}^\da D_{k\be}^\da D_{j\al} D_{k\be} \, \big]
\, , \label{e7} \eea
which must be followed by a projection on to the subspace of states satisfying
the constraint given in Eq. (\ref{b3}). The ground state of the system is the 
vacuum for the fermions, apart from a degeneracy of $2^m$ due to the presence
of $m$ zero energy modes. At low temperatures, the system will be described by
a dilute gas of fermions; as we saw above, the density of fermions is of order
$T$. For such a gas, the interaction energy per unit length between pairs of 
fermions belonging to different species is of order $T^3$, since the typical 
distance between two such fermions is of order $1/T$ and the interaction is 
inversely proportional to the square of the distance. On the other hand, the 
kinetic energy per unit length is proportional to $T^2$ as shown above. Hence,
the two-body interaction terms appearing in the second line of Eq. (\ref{e7}) 
can be ignored at low temperatures, and the corresponding Hamiltonian is well 
approximated by
\beq H ~=~ \sum_{\al=1}^m ~\sum_{u=0}^{N-1} ~E_u ~{\tilde D}_{u\al}^\da 
{\tilde D}_{u\al}, \label{e8} \eeq
where $E_u$ has the same form as in Eq. (\ref{e4}). Hence the low energy 
sector consists of $m$ species of non-interacting fermions each of which has 
the form of a massless Dirac fermion with only positive energy states. This 
is described by a CFT with central charge $c=m/2$. Using Eq. (\ref{e8}) and 
arguments similar to the ones given after Eqs. (\ref{e5}) and (\ref{e6}), 
one can show that the specific heat per unit length of this system goes as 
$\pi mT/(6v)$ at low temperatures.

It should be pointed out that for finite values of $N$, the partition function
of the model defined in Eq. (\ref{e2}) does {\it not} agree with that of $m$ 
species of non-interacting fermions, each with a dispersion relation given by 
Eq. (\ref{e4}). For instance, if we expand the two partitions functions, we 
obtain
\beq Z^{(m|1)}_{HS}(\tq) ~=~ 2^m ~[~ 1 ~+~ 2m ~\tq^{N-1} ~+~ m^2 ~\tq^{2N-2} ~
+~ m(m+1)~ \tq^{2N-4} ~+~ \dots ~] \label{e9} \eeq
from Eq. (\ref{b17}), and
\beq Z_1^m ~=~ 2^m ~[~ 1 ~+~ 2m ~\tq^{N-1} ~+~ m (2m-1) ~\tq^{2N-2} ~+~ 2m~ 
\tq^{2N-4} ~+~ \dots ~] \label{e10} \eeq
from Eq. (\ref{e5}). These two expressions do not agree at orders higher than
$\tq^{N-1}$, thereby showing the effect of two-particle interactions. However, 
since $\tq = e^{-\pi^2 /(N^2 T)}$, the difference between $\tq^{2N-2}$ and 
$\tq^{2N-4}$ becomes negligible in the limit 
$N \to \infty$; hence the total contribution from those terms becomes equal 
in Eqs. (\ref{e9}) and (\ref{e10}) since their coefficients add up to 
$m(2m+1)$ in both equations. Remarkably, we find that this kind of equality 
works up to all finite powers of $\tq^N$, although it fails for powers of 
$\tq^{N^2}$; the latter corresponds to contributions from high energy states 
whose energies are of order 1. If we now impose the condition that $T \ll 1$,
the terms of order $\tq^{N^2}$ go to zero. Motivated by this observation, we 
will explicitly prove in Sec. 6 that for $N \to \infty$ and $T \ll 1$, the 
partition function of the $SU(m|1)$ HS spin chain is identical to that 
of a model of $m$ species of non-interacting fermions.

The simple ground state structure of the $SU(m|1)$ HS spin chain implies that
the typical two-point equal-time correlation function in this model is trivial,
in contrast to the correlation function of the $SU(m|0)$ spin chain given in
Eq. (\ref{e1}). As discussed before Eq. (\ref{e7}), the $SU(m|1)$ HS spin 
chain is equivalent to a model in which there are $m$ species of fermionic 
spins and only species of bosonic spin which can be interpreted as a hole for 
the fermions. As in Eq. (\ref{e7}), we may define creation and annihilation 
operators for the $m$ species of fermions, $D_{j\al}^\da$ and $D_{j\al}$. We 
have seen that this system has $2^m$ ground states; for simplicity, let us 
first consider the ground state $|G \r$ in which there are no fermions and all
the sites are occupied by holes. Namely, $D_{j\al} |G \r = 0$ for all values 
of $j$ and $\al$. We then obtain the following two-point correlation function
\beq \l G | D_{j\al} D_{k \be}^\da | G \r ~=~ \de_{j k} ~\de_{\al \be}. 
\label{e11} \eeq
Even if we consider one of the other ground states in which there are $\ell$
fermions, where $1 \le \ell \le m$, the correlation function would have the
same form as in Eq. (\ref{e11}) in the thermodynamic limit $N \to \infty$.
This is because the factor of $1/\sqrt{N}$ in the definition of the Fourier 
transform as in Eq. (\ref{c13}) kills any contribution from the fermions 
if $\ell \ll N$. The simple form in Eq. (\ref{e11}) is in contrast to the 
correlation function for a system of non-interacting Dirac fermions in which 
the ground state has all one-particle states occupied up to some Fermi energy.
For such a ground state, the two-point correlation function defined in Eq. 
(\ref{e11}) typically falls off as $1/|j-k|$ in one dimension.

Finally, we would like to mention that the $SU(1|2)$ and $SU(1|m)$ spin chains
defined in Eq. (\ref{e2}) have been studied in Refs. \cite{arik} and 
\cite{thom} respectively. However, a chemical potential was implicitly 
introduced in those papers in order to consider ground states with a non-zero
filling of the fermions. Hence the ground states and excitations considered 
in Refs. \cite{arik,thom} differ from the ones that we have studied here. 

\no \section{Equivalence of the $SU(m|1)$ HS spin chain and $m$ species of 
non-interacting fermions}
\renewcommand{\theequation}{6.{\arabic{equation}}}
\setcounter{equation}{0}

In this section it will be shown that, in the thermodynamic limit and for low 
temperatures, the partition functions of the $SU(m|1)$ HS spin chain and a 
model of $m$ species of non-interacting fermions are equal to each other. 
As a by-product of this proof, we will derive a simple relation between the 
partition functions of the $SU(m|n)$ HS spin chain and the $SU(m|n)$ 
Polychronakos spin chain for any value of $n \ge 1$. While writing the above 
mentioned partition functions, we shall extensively use the notations 
introduced in Sec. 2. Let us begin by discussing the Polychronakos spin 
chain and its partition function.

In Ref. \cite{hika2}, the Hamiltonian of the $SU(m|n)$ supersymmetric 
Polychronakos spin chain is defined as 
\beq {\tilde H}^{(m|n)}_P ~=~ \frac{\pi^2}{N} ~\sum_{1\le j <k \le N} ~
\frac{1-{\tilde P}_{jk}^{(m|n)}}{(z_j - z_k)^2} , \label{f1} \eeq 
where the $z_j$'s are the roots of the $N$-th order Hermite polynomial $H_N 
(z)$. We have introduced a pre-factor of $\pi^2 /N$ in Eq. (\ref{f1}) for the 
following reason. While the distance between nearest neighbor sites is of order
$1/N$ in the HS spin chain ($\xi_j - \xi_{j+1}$ in Eq. (\ref{d1})), it is of 
order $1/\sqrt{N}$ in the Polychronakos spin chain ($z_j - z_{j+1}$ in Eq.
(\ref{f1})). The latter statement can be derived from the fact that the 
solution of the equation $d^2 H_N/dz^2 - 2z dH_N/dz + 2N H_N = 0$ is given by
$H_N \sim \cos (\sqrt{2N} z + N\pi /2)$ for $N \to \infty$ and $|z| \ll 
\sqrt{N}$; this region corresponds to sites near the middle of the chain. The 
zeros of this function have a spacing of $\pi /\sqrt{2N}$. Thus the Hamiltonian
in (\ref{f1}) takes the form
\beq {\tilde H}^{(m|n)}_P ~=~ 2 ~\sum_{j < k} ~\frac{1-{\tilde P}_{jk}^{
(m|n)}}{(j - k)^2} \label{f2} \eeq
for $j$ and $k$ lying close to $N/2$, as compared to the form given in Eq. 
(\ref{d2}). We thus see that the pre-factors in Eqs. (\ref{d1}) and (\ref{f1})
must differ by a factor of $N$ in order to ensure that the energy levels of 
the two Hamiltonians scale as the same power of $N$. We can then use the same 
variable $\tq = e^{-\pi^2 / (N^2 T)}$ when we compare the partition functions 
of the HS and Polychronakos spin chains.

According to Eq. (3.8) of Ref. \cite{hika2}, the partition function 
corresponding to the Hamiltonian in Eq. (\ref{f1}) can be written in the form
\beq {\tilde Z}^{(m|n)}_P(\tq) ~=~ \sum _{\mathbf{k} \in ~\mathcal{P}_N} ~
\tq^{ \frac{N^2(N-1)}{2} \, - \, N \sum \limits_{l=1}^{r-1} K_l} ~S_{\l k_1,
k_2,\dots, k_r \r}(x,y) \Bigr \rvert_{x=1,y=1} \, . \label{f3} \eeq
{}From this expression of the partition function, one obtains the eigenvalue 
of ${\tilde H}^{(m|n)}_P$ in (\ref{f1}) corresponding to the border strip 
$\l k_1,k_2,\dots,k_r \r$ as 
\beq {\tilde E}_{\l k_1,k_2,\dots,k_r \r} ~=~ \frac{\pi^2}{N} ~[~
\frac{N(N-1)}{2} ~-~ \sum_{l=1}^{r-1} ~ K_l ~]. \label{f4} \eeq
Let us now define the Hamiltonian of the $SU(m|n)$ Polychronakos spin chain 
in a slightly different form given by 
\beq H^{(m|n)}_P ~=~ \frac{\pi^2}{N} ~\sum_{1\le j <k \le N}
\frac{1+{\tilde P}_{jk}^{(m|n)}}{{(z_j - z_k )}^2}. \label{f5} \eeq 
Using Eqs. (\ref{f1}) and (\ref{f5}) along with an identity given by (see, 
for example, Ref. \cite{math})
\beq \sum_{1\le j <k \le N} \frac{1}{{(z_j-z_k)}^2} ~=~ \frac{N(N-1)}{4},
\label{f6} \eeq
it is easy to see that ${\tilde H}^{(m|n)}_P = \pi^2(N-1)/2 - H^{(m|n)}_P \, .$
Comparing this operator relation along with the eigenvalue relation (\ref{f4}),
we find the energy eigenvalue of $ H^{(m|n)}_P$ corresponding to the border 
strip $\l k_1,k_2,\dots,k_r \r$ to be
\beq E_{\l k_1,k_2,\dots,k_r \r} ~=~ \frac{\pi^2}{N} ~\sum_{l=1}^{r-1} ~K_l.
\label{f7} \eeq
In analogy with Eq. (\ref{f3}), we can write down the partition function
corresponding to the Hamiltonian in (\ref{f5}) as
\beq Z^{(m|n)}_P (\tq)~=~ \sum _{\mathbf{k} \in ~\mathcal{P}_N}
\tq^{\, N \sum \limits_{l=1}^{r-1} K_l} S_{\l k_1,k_2,\dots,
k_r \r}(x,y) \Bigr\rvert_{x=1,y=1}\, . \label{f8} \eeq

In Ref. \cite{basu5}, it has been shown that
\bea && \sum _{\mathbf{k} \in ~\mathcal{P}_N} ~\tq^{\, \sum 
\limits_{l=1}^{r-1} \mathcal{E}(K_l)}~ S_{\l k_1,k_2,\dots, k_r 
\r}(x,y) \Bigr \rvert_{x=1,y=1} \nn \\
&& ~~~~=~ \sum _{\mathbf{k} \in ~\mathcal{P}_N} ~\left( \prod^r_{i=1} 
d^{(m|n)}_{k_i} \right)~ \tq^{\, \sum\limits_{j=1}^{r-1}\mathcal{E}(K_j)} ~
\prod_{j=r+1}^N ~(1-\tq^{\mathcal{E}(K_j)}), \label{f9} \eea
where $\mathcal{E}(K_j)=K_j(N-K_j)$. 
We now observe that the proof of the relation in Eq. (\ref{f9}), as described 
in Sec. 3 of Ref. \cite{basu5}, remains valid if we choose $\mathcal{E}(K_j)=
K_jN$, instead of $\mathcal{E}(K_j)=K_j(N-K_j)$; in fact, the proof of this 
relation does not use any specific form of $\mathcal{E}(K_j)$. 
By using Eq. (\ref{f9}) for the case $\mathcal{E}(K_j)=K_jN$,
we can express the partition function $Z^{(m|n)}_P(\tq)$ in Eq. (\ref{f8}) as
\beq Z^{(m|n)}_P(\tq) ~=~ \left[ \prod_{j=1}^{N-1} (1-\tq^{\, jN})
\right]~ \sum_{r=1}^N ~\sum_{\substack{k_1+\dots +k_r=N, \\ k_j \ge 1} }~
\left( \prod^r_{i=1} d^{(m|n)}_{k_i} \right)~ \prod_{j=1}^{r-1} ~
\frac{\tq^{K_jN}}{1-\tq^{K_jN}} \, ,\label{f10} \eeq
where the summation over $\mathbf{k} \in ~\mathcal{P}_N$ is written explicitly
through its components. Eq. (\ref{f10}) is a new expression for the partition 
function of the $SU(m|n)$ Polychronakos spin chain; this expression is very 
similar in form to the partition function of the $SU(m|n)$ HS spin chain.

Let us now consider the limit $N \to \infty$ and $T<<1$, for which one can 
retain all terms with finite powers of $\tq^N$ and neglect terms of the order 
of $\tq^{N^2}$. Consequently, the dominant contribution on the right hand 
side of Eq. (\ref{f10}) comes from terms in which $k_1,\dots,k_{r-1}$
are of order $1$, and $k_r$ is close to $N$. For $k_r \sim N \to \infty$,
we have $d^{(m|n)}_{k_r} \to 2^m N^{n-1} /(n-1)!$. We thus obtain
\beq \lim_{N \to \infty} ~Z^{(m|n)}_P(\tq) ~=~ \frac{2^m 
N^{n-1}}{(n-1)!}~ \left[ \prod_{j=1}^{\infty}(1-\tq^{\, jN}) \right] ~
\sum_{k_1,\dots,k_l \ge 1} ~\prod^l_{j=1} ~\left( d^{(m|n)}_{k_j} ~
\frac{\tq^{K_jN}}{1-\tq^{K_jN}} \right). \label{f11} \eeq
The Hamiltonian ${\tilde H}^{(n|m)}_P$ in Eq. (\ref{f1}) can be related to 
$H^{(m|n)}_P$ in Eq. (\ref{f5}) through a unitary transformation described in
Eq. (\ref{b24}), $~U {\tilde H}^{(n|m)}_P U^\da ~=~ H^{(m|n)}_P.~$
Consequently, the corresponding partition functions satisfy the relation
\beq {\tilde Z}_P^{(n|m)}(\tq) ~=~ Z_P^{(m|n)}(\tq). \label{f12} \eeq

Using Eqs. (\ref{f11}) and (\ref{f12}) for the special case $n=1$, we obtain an
expression for the partition function of the $SU(1|m)$ Polychronakos spin chain
\beq \lim_{N \to \infty} ~ {\tilde Z}^{(1|m)}_P(\tq) ~=~2^m ~\left[ 
\prod_{j=1}^{\infty} (1-\tq^{\, jN}) \right] ~\sum_{k_1,\dots,k_l \ge 1} ~
\prod^l_{j=1} ~\left( d^{(m|1)}_{k_j} ~\frac{\tq^{K_jN}}{1-\tq^{K_jN}} \right).
\label{f13} \eeq
On the other hand, it is shown in Ref. \cite{hika2} that
\beq \lim_{N \to \infty} {\tilde Z}_P^{(1|m)}(\tq) ~=~ \prod_{j=0}^\infty ~(1+
\tq^{\, jN})^m. \label{f14} \eeq
Comparing the right hand sides of Eqs. (\ref{f13}) and (\ref{f14}), we get
\beq \left[ \prod_{j=1}^{\infty}(1-\tq^{\, jN}) \right] ~\sum_{k_1,\dots,
k_l \ge 1} ~\prod^l_{j=1}~ \left( d^{(m|1)}_{k_j} ~\frac{\tq^{K_jN}}{1-
\tq^{K_jN}} \right) ~=~\prod_{j=1}^\infty ~(1+\tq^{\, jN})^m . \label{f15} \eeq
Squaring both sides of this equation and multiplying by $2^m$, we obtain
\beq 2^m ~\left[ \prod_{j=1}^{\infty}(1-\tq^{\, jN})^2 \right] ~\left[ ~
\sum_{k_1,\dots,k_l \ge 1} ~\prod^l_{j=1}~ \left( d^{(m|1)}_{k_j} ~
\frac{\tq^{K_jN}}{1-\tq^{K_jN}} \right)~ \right]^2 ~= ~2^m ~
\prod_{j=1}^\infty ~ (1+\tq^{\, jN})^{2m} . \label{f16} \eeq

We will now prove the equivalence of the partition functions of the $SU(m|1)$ 
HS spin chain and $m$ species of non-interacting fermions in the limits $N 
\to \infty$ and $T \ll 1$, by showing that the left hand side of (\ref{f16}) 
is equal to the partition function of the $SU(m|1)$ HS spin chain, while the 
right hand side of (\ref{f16}) is the partition function of $m$ species of 
non-interacting fermions. For $N \to \infty$ and $T \ll 1$, 
the partition function of one fermion given in Eq. (\ref{e5}) only gets 
contributions from values of $u$ close to either 0 or $N$. The term with $u=0$
contributes a factor of 2, while the terms with $u$ non-zero and close to 0 
and $u$ close to $N$ each contribute $\prod_{j=1}^\infty ~(1+\tq^{jN})$.
Putting these together, we see that the partition function of $m$ 
non-interacting fermions is equal to the right hand side of Eq. (\ref{f16}). 
Similarly, for $N \to \infty$ and $T \ll 1$, we find that the only terms which
contribute in Eq. (\ref{b15}) are those partitions ${\bf k} =\{k_1, k_2, 
\dots, k_s, \dots, k_r \}$ in which each of the $k_i$'s is of order 1 except 
for one, say, $k_s$ which is close to $N$. In the limit $k_s \sim N \to 
\infty$, we have $d^{(m|1)}_{k_s} \to 2^m $. Therefore, we can write 
$d^{(m|1)} ({\bf k})$ in Eq. (\ref{b11}) as
\beq d^{(m|1)} ({\bf k}) ~=~ \left( ~2^{m-1}\prod_{a=1}^{s-1}~ d^{(m|1)}(k_a)~
\right) ~\left(2^{m-1} ~\prod_{b=s+1}^r ~d^{(m|1)} (k_b)~\right). 
\label{f17} \eeq
Further, for the above mentioned partitions, the value of partial sums $K_1, 
K_2, \dots , K_{s-1}$ are close to $0$ and the value of partial sums $K_s, 
K_{s+1}, \dots , K_{r-1}$ are close to $N$. One can approximate $\tq^{K_j 
(N-K_j)}$ by $\tq^{K_j N}$ if $K_j$ is close to $0$ , and by $\tq^{(N-K_j) N}$
if $K_j$ is close to $N$ in Eq. (\ref{b15}). Combining this result along with 
the form of $d^{(m|1)} ({\bf k})$ given in Eq. (\ref{f17}), we find that the 
contributions of the terms with $K_j$ close to 0 and the terms with $K_j$ 
close to $N$ have the same form for various partitions $\bf k$ in Eq. 
(\ref{b15}); each of them is given by
\beq 2^{m-1}\left[ \prod_{j=1}^{\infty}(1-\tq^{jN}) \right] ~\sum_{k_1,\dots, 
k_l \ge 1}~\prod^l_{j=1}~ \left( d^{(m|1)}_{k_j} ~\frac{\tq^{K_jN}}{1- 
\tq^{K_jN}} \right), \label{f18} \eeq
where all the $k_j$'s and $K_j$'s are now of order 1. The partition function 
of the $SU(m|1)$ HS spin chain is evidently obtained by taking the square of 
the expression in Eq. (\ref{f18}). Thus we find that, in the limit $N \to 
\infty$ and $T \ll 1$, the partition function of the $SU(m|1)$ HS spin chain 
coincides with left hand side of Eq. (\ref{f16}).

Finally, let us note that the derivation of Eqs. (\ref{f17}) and (\ref{f18}) 
given above for $n=1$ can be generalized easily to any value of $n \ge 1$. 
Combining this with Eq. (\ref{f11}), we see that for $N \to \infty$ and 
$T \ll 1$, the partition functions of the $SU(m|n)$ HS and Polychronakos 
spin chains are related as
\beq \frac{2^m N^{n-1}}{(n-1)!} ~Z_{HS}^{(m|n)} (\tq) ~=~ \left[ Z_P^{(m|n)} 
(\tq) \right]^2. \label{f19} \eeq

\no \section{Conclusions}

In this paper, we have used the exact partition function of the $SU(m|n)$ HS
spin chain to find its complete spectrum, including the degeneracy of all 
energy levels, in terms of the motif representations. We have also obtained 
the momentum eigenvalue associated with different motifs.
We have then studied the ground state and
low energy excitations of the $SU(m|n)$ HS spin chain with $N$ sites, for 
various values of $m$, $n$ and $N$. In the thermodynamic limit $N \to \infty$,
the low energy, low momentum spectrum is always found to have a linear relation
between the energy and momentum, with the velocity being independent of $m$ and
$n$. The $SU(m|0)$ spin chain has some low energy, high momentum excitations 
which may be related to the algebraic long-range order of the system.

In the thermodynamic limit, the ground state degeneracy remains finite only 
for the $SU(m|0)$ and $SU(m|1)$ HS spin chains. Hence the low energy 
excitations of only these spin chains can possibly be described by conformal 
field theories. The $SU(m|0)$ spin chain is known to be described by the 
$SU(m)_1$ WZNW CFT with central charge $m-1$. We have derived exact expressions
for the partition function of the $SU(1|1)$ spin chain for any value of $N$, 
and of the $SU(m|1)$ spin chain for $m \ge 2$ in the limit $N \to \infty$ and 
the temperature $T \ll 1$. We have shown that for all $m \ge 1$, the
low temperature properties of the $SU(m|1)$ HS spin chain are the same as 
those of a model of $m$ non-interacting Dirac fermions, each of which has only
positive energy states. Such a theory has central charge $m/2$.

Finally, we have shown that in the thermodynamic limit and at low temperatures,
the partition function of the $SU(m|n)$ HS spin chain is related to the square
of the partition function of the $SU(m|n)$ Polychronakos spin chain for $n \ge
1$.

\vskip 1 cm
\no {\bf Acknowledgments}
\vskip .4 cm

We thank Sriram Shastry for a discussion of the spin-1/2 chain. D.S. thanks 
DST, India for financial support under projects SR/FST/PSI-022/2000 and 
SR/S2/CMP-27/2006.

\vskip .5 cm

\section*{\begin{large}Appendix A. \end{large}
\begin{normalsize}
Equivalence of the $SU(1|1)$ HS spin chain and one species of non-interacting
fermions \end{normalsize}}
\renewcommand{\theequation}{A\arabic{equation}}
\setcounter{equation}{0}

By using Eq. (\ref{b12}) we find that $d_{k_i}^{(1|1)}=2$ for any value of 
$k_i$. Substituting this value of $d_{k_i}^{(1|1)}$ in Eq. (\ref{b20}), and 
writing the corresponding summation variable $\mathbf{l} ~(\in ~\mathcal{P}_r)$
through its components, we obtain
\bea S_{\l k_1,k_2,\dots,k_r \r}(x,y)\rvert_{x=1,y=1}
&=& \sum_{s=1}^r ~\sum_{\substack{ \ell_1 +\ell_2 +\dots + \ell_s =N \\
\ell_s \ge 1}} ~(-1)^{r-s} ~\prod_{i=1}^s ~2 \nn \\
&=& \sum_{s=1}^r (-1)^{r-s} ~2^s ~\sum_{\substack{ \ell_1 +\ell_2 +\dots + 
\ell_s =N \\ \ell_s \ge 1}} ~1 \nn \\
&=& \sum_{s=1}^r(-1)^{r-s} ~2^s ~f(r,s). \eea
Here $f(r,s)$ indicates the number of possible ways of partitioning $r$ into 
length $s$, taking care of ordering. Clearly, this problem is equivalent to 
the problem of distributing $r$ identical balls amongst $s$ identical boxes, 
where each box contains at least one ball. After putting one ball in each box,
there will be $r-s$ balls remaining, which can be distributed freely amongst 
the $s$ different boxes. This problem is the same as distributing $r-s$ bosons
amongst $s$ states. Therefore the number of different distributions is 
\beq f(r,s) ~=~ \frac{(r-s+s-1)!}{(s-1)!(r-j)!} ~=~ {}^{r-1}C_{s-1}. \eeq
Substituting this in Eq. (A1), we obtain
\bea S_{ \l k_1,k_2,\dots,k_r \r}(x,y)\rvert_{x=1,y=1}
&=& 2 \, \sum_{s=1}^r \, {}^{r-1}C_{s-1} \, 2^{s-1} \, (-1)^{r-s} \nn \\
&=& 2 ~\sum_{s=0}^{r-1}\, {}^{r-1}C_s\, 2^s \, (-1)^{r-s+1} \nn \\ 
&=& 2 \, (2-1)^{r-1} ~=~ 2. \eea
Substituting this value of $S_{ \l k_1,k_2,\dots,k_r \r}(x,y)\rvert_{x=1,y=1}$
in Eq. (\ref{b18}), we find that
\bea Z_{HS}^{(1|1)} (\tq)
&=& \sum_{r=1}^N ~\sum_{\substack{k_1+\dots +k_r=N, \\ k_j \ge 1} }
\, 2 \, \tq^{\sum\limits_{j=1}^{r-1} \mathcal{E}(K_j)} \nn \\
&=& 2 ~\sum_{r=1}^N ~\sum_{1 \le K_1 < K_2< \dots < K_{r-1} \le N-1}~
\tq^{\sum\limits_{j=1}^{r-1} \mathcal{E}(K_j)} \, ,\eea
where $\mathcal{E}(K_j)=K_j (N-K_j)$. 

To compare $Z_{HS}^{(1|1)}(\tq)$ with the partition function of one species
of non-interacting fermions, let us expand the fermion partition function in 
Eq. (\ref{e5}) as follows:
\bea 2 ~\prod_{j=1}^{N-1} ~(1+\tq^{\mathcal{E}(j)}) 
&=& 2 ~(1+\tq^{\mathcal{E}(1)}) ~(1+\tq^{\mathcal{E}(2)}) \dots 
(1+\tq^{\mathcal{E}(N-1)}) \nn \\
&=& 2 ~\left[ 1 ~+~ \sum_{l_1=1}^{N-1} ~\tq^{\mathcal{E}(l_1)} ~+~ 
\sum_{1\le l_1 <l_2 \le N-1} ~\tq^{ \mathcal{E}(l_1)+ \mathcal{E}(l_2)}~
+~ \dots \right] \nn \\
&=& 2 ~\sum_{s=0}^{N-1} ~\sum_{1\le l_1 <l_2 \dots <l_s \le N-1}~ 
\tq^{\sum_{j=1}^s \mathcal{E}(l_j)}. \eea
Comparing Eqs. (A4) and (A5), we find complete equivalence between the 
partition function of the $SU(1|1)$ HS spin chain and that of one species of 
non-interacting fermions for any value of $N$ and $T$.


\newpage
\begin{figure}[h]
\centering 
\hskip -.25 cm 
\includegraphics{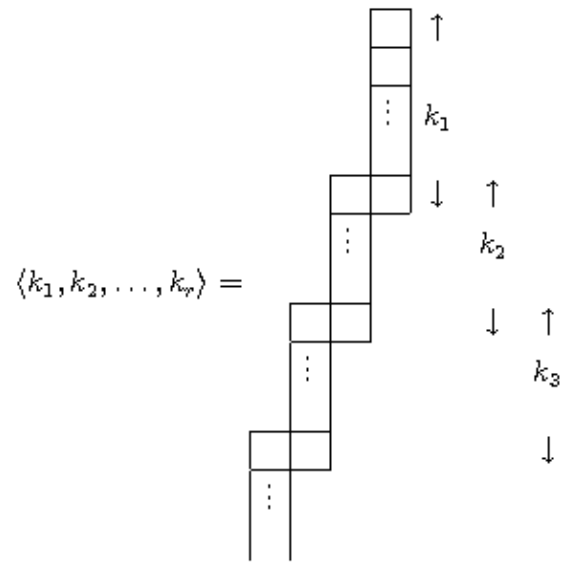}
\caption{Shape of the border strip $\l k_1, k_2, \dots , k_r \r$.}
\end{figure}

\newpage
\begin{figure}[h]
\centering 
\hskip -.25 cm 
\includegraphics[]{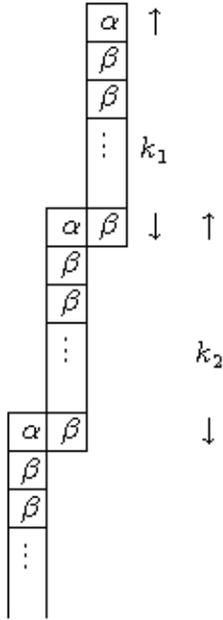}
\caption{Form of an allowed tableau corresponding to the border strip $\l k_1,
k_2, \dots ,k_r \r$ (with arbitrary values of $k_i$) occurring in the Fock 
space of the $SU(m|n)$ supersymmetric HS spin chain. Here $\alpha$ is any 
number within the set $ \{ 1,2, \dots , m \}$, and $\beta$ is any number 
within the set $\{ m+1, m+2, \dots , m+n \}$.} 
\end{figure}

\end{document}